\documentclass[10pt]{iopart}

\usepackage{iopams}
\usepackage{multirow}
\usepackage{booktabs}
\expandafter\let\csname equation*\endcsname\relax
\expandafter\let\csname endequation*\endcsname\relax
\usepackage{amsmath}
\usepackage{amssymb}
\usepackage{bm}
\usepackage{tabularx}
\usepackage{cite}
\usepackage{soul}

\usepackage[pdfborder={0 0 0}]{hyperref}
\usepackage{graphicx}
\usepackage{grffile}

\graphicspath{{./figs/}}

\usepackage[usenames,dvipsnames]{xcolor}

\begin{document}

\title[Kagome materials $A$V$_3$Sb$_5$ ($A$=K,Rb,Cs): pairing symmetry and pressure-tuning studies]{Kagome materials $A$V$_3$Sb$_5$ ($A$=K,Rb,Cs): pairing symmetry and pressure-tuning studies}

\newcommand{\orcid}[1]{\href{https://orcid.org/#1}{\raisebox{2.5pt}{\includegraphics[width=7pt]{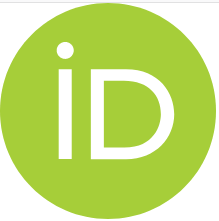}}}}

\author{
   Yuwei Zhou$^1$, Ge Ye$^1$, Shuaishuai Luo$^1$, Yu Song$^{1,}$$^*$, Xin Lu$^{1,}$$^{2,}$$^*$, Huiqiu Yuan$^{1,}$$^{2,}$$^{3,}$$^*$
}
\address{$^1$ Center for Correlated Matter and School of Physics, Zhejiang University, 310058 Hangzhou, China\par
$^2$ Zhejiang Province Key Laboratory of Quantum Technology and Device, School of Physics, Zhejiang University, Hangzhou 310058, China\par
$^3$ State Key Laboratory of Silicon Materials, Zhejiang University, Hangzhou 310058, China}

\ead{yusong\_phys@zju.edu.cn($^*$corresponding author)}
\ead{xinluphy@zju.edu.cn}
\ead{hqyuan@zju.edu.cn}

\ioptwocol

\begin{abstract}
The vanadium-based kagome metals $A$V$_3$Sb$_5$ ($A$ = K, Rb, and Cs) host a superconducting ground state that coexists with an unconventional charge density wave (CDW). The CDW state exhibits experimental signatures of chirality, electronic nematicity, and time-reversal-symmetry-breaking, raising the questions whether the superconductivity (SC) in  $A$V$_3$Sb$_5$ may also be unconventional, how SC interplays with CDW, and how the two orders evolve upon tuning. This article reviews studies of the superconducting pairing symmetry, and the tuning of SC and CDW in the $A$V$_3$Sb$_5$ compounds. Various experimental techniques consistently find that CsV$_3$Sb$_5$ exhibits nodeless SC, which remains robust regardless whether the CDW is present. Under hydrostatic pressure, SC in $A$V$_3$Sb$_5$ becomes enhanced as the CDW is gradually suppressed, revealing a competition between the two orders. In CsV$_3$Sb$_5$, a new CDW state emerges under pressure that competes more strongly with SC relative to the CDW at ambient pressure, and results in two superconducting domes that coexist with CDW. After the CDW in $A$V$_3$Sb$_5$ is fully suppressed with hydrostatic pressure, a further increase in pressure leads to a nonmonotonic evolution of the superconducting transition temperature driven by lattice modulations. Thickness is shown to be a powerful tuning parameter in $A$V$_3$Sb$_5$ thin flakes, revealing the evolution of CDW and SC upon dimensional reduction, and can be combined with hydrostatic pressure to shed light on the interplay between SC and CDW. Based on results reviewed in this article, we discuss outstanding issues to be addressed in the $A$V$_3$Sb$_5$ systems. 
\end{abstract}

\tableofcontents

\section{Introduction}

The kagome lattice consists of corner-sharing triangles, which leads to intrinsic frustration, and thus insulating kagome materials have been studied extensively as platforms for quantum magnetism and quantum spin liquids \cite{norman2016colloquium,zhou2017quantum,balents2010spin}. In kagome metals, the unique geometry of the kagome lattice leads to Dirac points, flat bands, and van Hove singularities (vHs) in the electronic structure, promoting topologically non-trivial and correlated states of matter \cite{yin2022topological}. The recently discovered vanadium-based kagome $A$V$_3$Sb$_5$ ($A$=K, Rb, Cs) materials exhibit a superconducting ground state that emerges from a highly unusual charge-density-wave (CDW) \cite{ortiz2019new,ortiz2020cs,ortiz2021superconductivity,yin2021superconductivity}, and triggered a flurry of research activities on their physical properties.\par

\begin{figure*}[htbp]
  \centering
  \includegraphics[width=\textwidth]{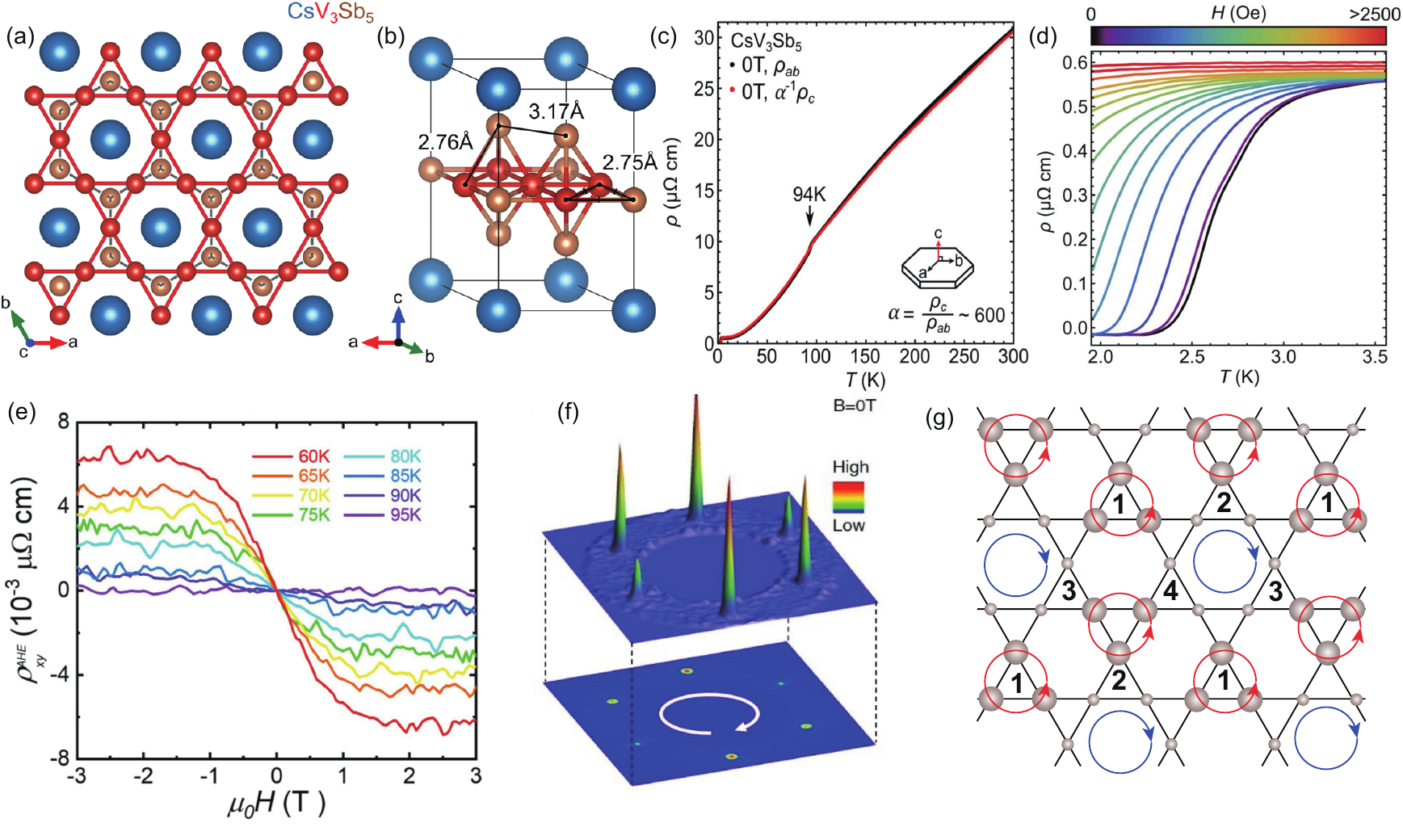}
  \caption{
  The crystal structure and resistivity of CsV$_3$Sb$_5$\cite{ortiz2020cs} include a schematic of the ab-plane(a), the three-dimensional crystal structure (b) with V atoms (colored in red) forming an ideal kagome lattice, and resistivity measurements (c)-(d) for CsV$_3$Sb$_5$, which allow $T_{\rm CDW}$ and $T_{\rm c}$ to be determined. (e) Anomalous Hall effect in CsV$_3$Sb$_5$ \cite{yu2021concurrence}, which appears as the CDW develops. (f) Chiral CDW in KV$_3$Sb$_5$ probed by scanning tunneling microscopy \cite{jiang2021unconventional}. (g) Schematic of the chiral flux phase \cite{feng2021chiral}.
  }
    \label{fig:fig1}
\end{figure*}

$A$V$_3$Sb$_5$ crystallizes in a layered structure (P6/mmm space group), as shown in Figs.~\ref{fig:fig1}(a) and (b). Triangular layers of $A$ atoms separate layers of V-Sb atoms, with the V atoms forming an ideal two-dimensional kagome lattice in the $ab$-plane. The layered nature of $A$V$_3$Sb$_5$ compounds plays a key role in their physical properties, including the highly anisotropic resistivity and the lack of apparent dispersion along $k_z$ in the electronic structure measured by angle-resolved photoemission spectroscopy (ARPES) \cite{ortiz2021fermi}. Density functional theory (DFT) calculations and ARPES measurements reveal multiple Dirac points and topologically non-trivial surface states close to the Fermi level, which identifies the P6/mmm phase of $A$V$_3$Sb$_5$ as a $Z_2$ topological metal \cite{ortiz2019new,ortiz2020cs,lou2022charge,cho2021emergence,nakayama2021multiple,liu2021charge,luo2022electronic,kang2022twofold,hu2022topological}. Upon cooling, all three member compounds of the $A$V$_3$Sb$_5$ family exhibit a CDW transition which expands the unit cell by $2\times2$ in the $ab$-plane (CDW transition temperatures $T_{\rm CDW}=78$~K, 103~K, 94~K for $A$=K, Rb, Cs, respectively), and a superconducting phase emerges from the CDW phase (superconducting transition temperatures $T_{\rm c}=0.9$~K, 0.9~K, 2.5~K for $A$=K, Rb, Cs) \cite{ortiz2020cs,ortiz2021superconductivity,yin2021superconductivity}. ARPES \cite{hu2022coexistence}, X-ray diffraction (XRD)\cite{ortiz2021fermi,stahl2022temperature,xiao2023coexistence,kautzsch2023structural}, and nuclear magnetic resonance (NMR) measurements \cite{luo2022posiible} have further indicated the coexistence of 2$\times$2$\times$2(tri-hexagonal distortions) and 2$\times$2$\times$4(with both tri-hexagonal and star-of-David distortions \cite{ortiz2021fermi} or just tri-hexagonal distortions \cite{xiao2023coexistence}) CDW orders in CsV$_3$Sb$_5$. In contrast, the CDW in KV$_3$Sb$_5$ and RbV$_3$Sb$_5$ are 2$\times$2$\times$2, without signatures of a coexisting 2$\times$2$\times$4 CDW \cite{ frassineti2023microscopic}. 

Studies of the CDW state in $A$V$_3$Sb$_5$ revealed a number of unusual properties, which show that the CDW not only breaks translational symmetry, but also shows additional symmetry-breaking. Using scanning tunneling microscopy (STM), it was found the Fourier weights of the 2$\times$2 CDW peaks can be switched between clockwise to anti-clockwise via an external magnetic field, which suggest the CDW state to be chiral \cite{jiang2021unconventional} (Fig.~\ref{fig:fig1}(f)) and breaks time-reversal symmetry \cite{shumiya2021intrinsic,jiang2021unconventional,wang2021electronic}. Although there are no local moments or long-range magnetic order in $A$V$_3$Sb$_5$ \cite{mielke2022time,khasanov2022time,yu2021evidence,hu2022time}, anomalous Hall and anomalous Nernst effects (Fig.~\ref{fig:fig1}(e)) appear below $T_{\rm CDW}$, which can also be interpreted as evidence for time-reversal symmetry-breaking (TRSB) \cite{yang2020giant,yu2021concurrence,zhou2022anomalous,chen2022anomalous}. The onset temperatures of TRSB remain under debate \cite{denner2021analysis,feng2021low,hu2022time,wu2022simultaneous,xu2022three,yu2021evidence,asaba2023evidence}. Magneto-optical Kerr effect\cite{hu2022time}, optical polarization rotation measurements\cite{wu2022simultaneous} and scanning birefringence microscopy \cite{xu2022three} find that TRSB occurs at $T_{\rm CDW}$. On the other hand, zero-field muon spin relaxation measurements ($\mu$SR) \cite{yu2021evidence} finds TRSB appear at a temperature well below $T_{\rm CDW}$, and magnetic torque measurements \cite{asaba2023evidence} reveal hints of TRSB at temperatures above $T_{\rm CDW}$. A possible origin for the TRSB of the 2$\times$2 CDW is the chiral flux phase \cite{feng2021low,feng2021chiral,denner2021analysis,lin2021complex}, where two types of current flux which respectively form honeycomb and triangular lattices, loop with opposite directions and result in a net magnetization (Fig.~\ref{fig:fig1}(g)). 

In addition to TRSB, the CDW in $A$V$_3$Sb$_5$ system also leads other consequences, with reports of CDW-induced band renormalization and energy gaps\cite{liu2021charge}, changes in the local magnetic field well inside the CDW state \cite{khasanov2022time}, and an electronic nematic transition breaking six-fold rotational symmetry well inside the CDW state \cite{nie2022charge}. The superconducting state which emerges from such an unusual CDW state, is also found to exhibit unconventional features, such as a pair-density wave \cite{chen2021roton}, two-fold rotational symmetry \cite{xiang2021twofold,wang2024two}, and Majorana zero modes in vortex cores \cite{liang2021three}. Given these unconventional behaviors of the SC and that it emerges from a highly unusual CDW, the pairing symmetry and how SC interplays with CDW are key questions in understanding the physics of $A$V$_3$Sb$_5$ compounds.

Although research on $A$V$_3$Sb$_5$ is still at a young age, concerted research efforts have uncovered many fascinating behaviors and underlying mechanisms, which have been summarized in several excellent reviews \cite{jiang2023kagome,chen2022superconductivity,neupert2022charge,mi2023electrical,yin2022topological,nguyen2022electronic,wilson2024kagome}. This article, which partially overlaps with an earlier Chinese review\cite{zhang2023superconducting}, focuses on a few selected topics in the research of $A$V$_3$Sb$_5$, including (1) the superconducting order parameter; (2) the interplay between SC and CDW, revealed upon pressure- and thickness-tuning; and (3) the evolution of SC with pressure after the CDW is fully suppressed. These results are an integral part of the $A$V$_3$Sb$_5$ systems, and provide the basis for further in-depth studies. 
 
\section{Superconducting order parameter in $A$V$_3$Sb$_5$ at ambient pressure}

When the kagome lattice is close to van Hove filling, unconventional SC with $f$-wave pairing \cite{wu_2021} or $d+id$-wave pairing \cite{kiesel_2012} may arise. Given the unusual CDW state from which SC emerges, it is important to understand the superconducting pairing symmetry in $A$V$_3$Sb$_5$, particularly whether the SC is conventional with a sign-preserving order parameter or unconventional with a sign-changing order parameter. The character of the superconducting order parameter also serves as the basis for understanding the pairing mechanism, providing strong constraints on theoretical models. 

Early studies of K$_{1-x}$V$_3$Sb$_5$ Josephson junctions provided hints of spin-triplet SC \cite{wang_2020} , and thermal conductivity measurements of CsV$_3$Sb$_5$ suggested nodal SC \cite{zhao2021nodal}. These studies spurred examinations of the superconducting pairing symmetry in $A$V$_3$Sb$_5$, and in particular CsV$_3$Sb$_5$, using a variety of experimental techniques, including tunnel diode oscillator (TDO) \cite{duan2021nodeless,roppongi2023bulk}, $\mu$SR\cite{shan2022muon,gupta2022microscopic,gupta2022two}, STM\cite{xu2021multiband,liang2021three}, point-contact spectroscopy (PCS)\cite{yin2021strain,he2022strong}, NMR\cite{mu2021s}, and ARPES \cite{zhong2023nodeless,mine2024direct}. These studies collectively find that the SC in CsV$_3$Sb$_5$ is spin-singlet and nodeless, consistent with a conventional phonon-driven pairing mechanism. 

Measurements of the magnetic penetration depth $\lambda(T)$,  typically carried out using TDO and $\mu$SR, provides a straightforward and sensitive probe of the pairing symmetry in superconductors. Transverse field $\mu$SR (TF-$\mu$SR) allows for quantitative measurement of $\lambda(T)$, although it is often challenging to obtain high-precision data due to its time-consuming nature. Moreover, these measurements are typically carried out under a sizable externally field, which in some circumstances could affect the measurement outcomes. On the other hand, TDO is a highly efficient self-resonating technique \cite{van_1975} that utilizes the negative resistance of tunnel diodes to probe $\lambda(T)$ with high precision, and only requires an extremely small magnetic field. However, TDO only measures the relative change of the magnetic penetration depth with temperature ($\Delta\lambda(T)=\lambda(T)-\lambda(0)$), and is unable to directly probe the absolute value of $\lambda(T)$. Therefore,  TDO and $\mu$SR are complementary techniques in probing materials' superconducting order parameters via measurements of the magnetic penetration depth $\lambda(T)$.

\begin{figure}[t]
  \centering
  \includegraphics[width=\columnwidth]{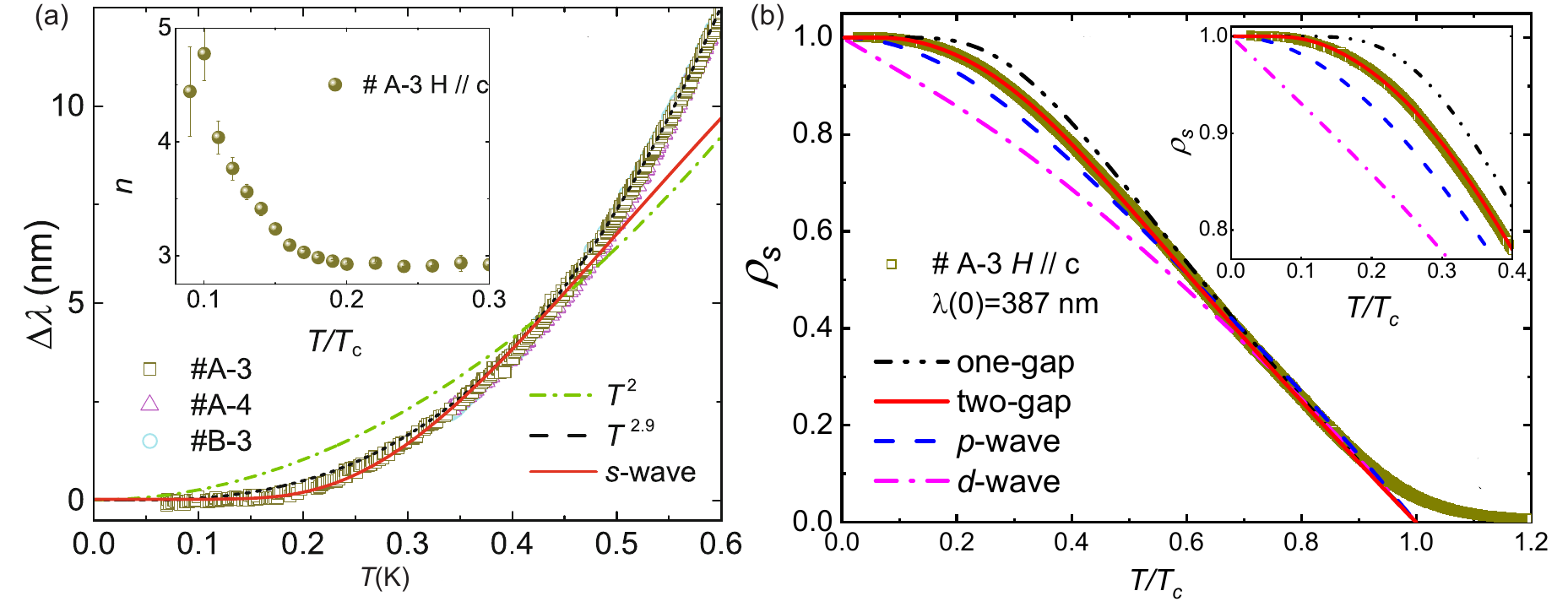}
  \caption{
   (a) The  change of the magnetic penetration depth $\Delta\lambda(T)$, measured for three CsV$_3$Sb$_5$ samples\cite{duan2021nodeless}. The red solid line represents a fit to the $s$-wave model. The green and black dashed lines respectively represent fits to a $T^2$ and a $T^n$ temperature dependence. The inset shows the exponent $n$ obtained by fitting $\Delta\lambda(T)$ to $T^n$ from 0.07 K up to different values of $T/T_c$.
    (b) The temperature dependence of the normalized superfluid density $\rho_{\rm s}(T)$ in CsV$_3$Sb$_5$. The red solid line represents a fit to the two-band $s$-wave model. The black, blue and purple dashed lines represent fits to single-gap $s$-wave, $p$-wave, and $d$-wave model, respectively. The inset zooms in at low temperatures.
    }
    \label{fig:fig3_4_2}
 \end{figure}

The magnetic field penetration depth $\Delta\lambda(T)$ of CsV$_3$Sb$_5$ single-crystals measured via TDO are shown in Fig.~\ref{fig:fig3_4_2}  \cite{duan2021nodeless}. The results indicate that $\Delta\lambda(T)$ at low temperatures conforms to an exponential temperature dependence (Fig.~\ref{fig:fig3_4_2} (a)), providing direct evidence for the absence of nodes in the superconducting gap structure of CsV$_3$Sb$_5$. This conclusion is further corroborated by fitting the data to a power law $\sim T^n$, with the exponent $n$ obtained from the fitting (the inset of Fig.~\ref{fig:fig3_4_2}(a)) consistently exceeding 2, which is the upper limit in the presence of nodes. 

The magnetic penetration depth $\Delta\lambda(T)$ of CsV$_3$Sb$_5$ at low temperatures can be captured by a single-gap $s$-wave model and yields $\Delta(0)=0.59k_{\rm B}T_{\rm c}$ (Fig.~\ref{fig:fig3_4_2}(a)), which is much smaller than the weak coupling limit.  This suggests the presence of multiple superconducting energy gaps in CsV$_3$Sb$_5$, and is consistent with the analysis of the normalized superfluid density $\rho_{\rm s}(T)=\frac{\lambda(0)^2}{[\Delta\lambda(T)+\lambda(0)]^2}$, which can be described by a two-gap $s$-wave model (Fig.~\ref{fig:fig3_4_2}(b)). The conclusion of two-gap nodeless SC remains robust upon varying $\lambda(0)$ by $\approx\pm20\%$. 

\begin{figure*}[htbp]
  \centering
  \includegraphics[width=\textwidth]{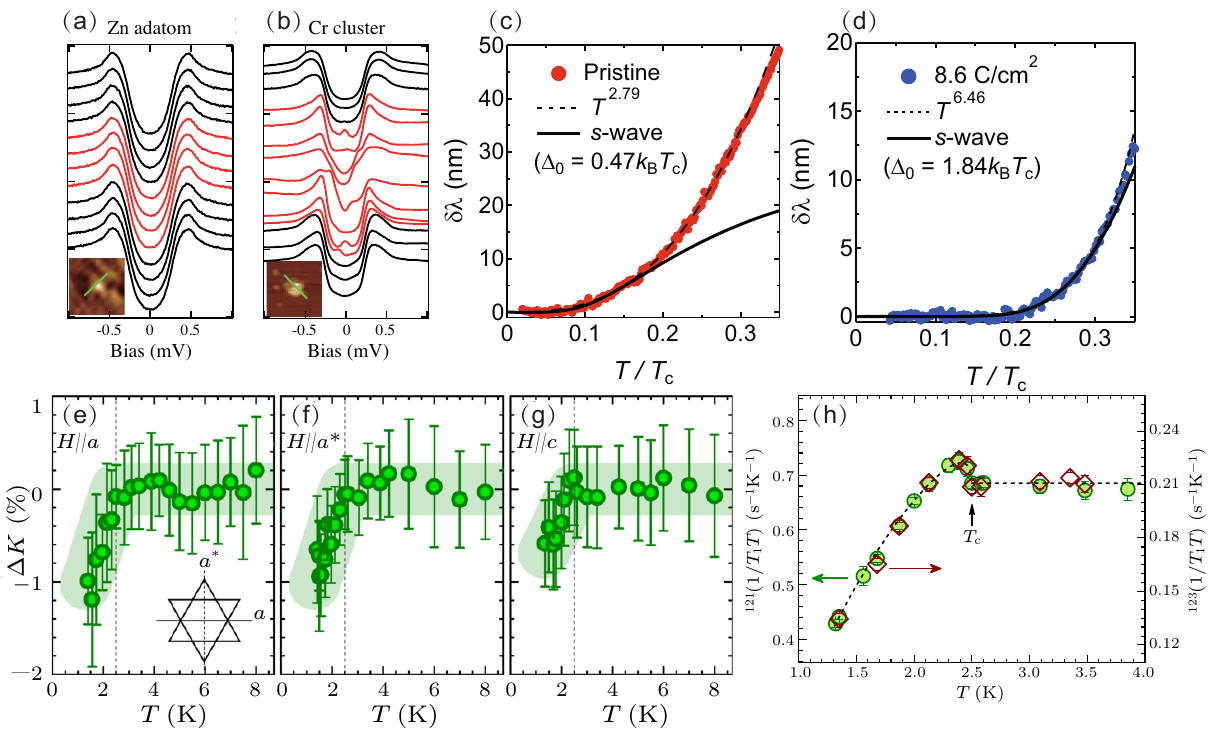}
  \caption{
   Evidence of spin-singlet and sign-preserving SC in CsV$_3$Sb$_5$.
  (a)-(b) SC of CsV$_3$Sb$_5$ is strongly suppressed at magnetic Cr clusters, whereas a negligible effect is seen at nonmagnetic Zn impurities \cite{xu2021multiband}.
  (c)-(d) Temperature dependence of the magnetic penetration depth change $\Delta\lambda(T)$, for the pristine and the 8.6 C/cm$^{2}$ irradiated samples\cite{roppongi2023bulk}. The black solid and dashed lines are fitting curves to the fully-gapped $s$-wave model and a $T^n$ behavior.      
  (e)-(g) The decrease of the Knight shift $\Delta K$ below $T_{\rm c}$, for fields along various directions\cite{mu2021s}. (h)  Observation of a Hebel–Slichter coherence peak just below $T_{\rm c}$ in measurements of the spin-lattice relaxation rate $1/T_1T$ \cite{mu2021s}.  
 }
    \label{fig:fig_nosign}
 \end{figure*}

 ARPES allows the superconducting gaps to be directly probed with momentum resolution, although the small gap sizes of $A$V$_3$Sb$_5$ make such measurements technically challenging. To overcome this difficulty, chemical substitution was used to weaken or suppress the CDW and enhance SC in CsV$_3$Sb$_5$, and the superconducting gaps in Cs(V$_{0.93}$Nb$_{0.07}$)$_3$Sb$_5$ (with weakened CDW) and Cs(V$_{0.86}$Nb$_{0.14}$)$_3$Sb$_5$ (without CDW) were directly measured using ARPES (Fig.~\ref{fig:fig_doping2}(d)-(e)), revealing isotropic gaps on all Fermi surfaces regardless of  whether SC coexists with CDW or not, and suggests nodeless SC to be a robust feature in CsV$_3$Sb$_5$ and derived compounds \cite{zhong2023nodeless}. A more recent ARPES study of pure CsV$_3$Sb$_5$ confirms the superconducting gaps to be nodeless, but finds a strong in-plane anisotropy of the gap derived from V-3$d$ orbitals, in contrast to the isotropic gaps derived from Sb-5$p$ orbitals \cite{mine2024direct}. 
 
Given that $A$V$_3$Sb$_5$ contains multiple Fermi surfaces, it is important to determine whether there are sign changes of the superconducting order parameter between different Fermi surfaces. To this end, systematic impurity effect studies were carried out for CsV$_3$Sb$_5$, which consistently find sign-preserving SC \cite{xu2021multiband,roppongi2023bulk,zhang2023nodeless}. The impact of local impurities on SC is strongly influenced by the pairing symmetry, as well as by whether the impurities are magnetic \cite{balatsky_2006}. In the case of sign-preserving SC, magnetic impurities strongly disrupt Cooper pairing, and leads to in-gap bound states\cite{anderson_1959}.  For sign-changing pairing such as $d$-wave and $s\pm$-wave, nonmagnetic impurities also generate in-gap states and inhibit superconductivity\cite{prozorov_2006}. In STM measurements of CsV$_3$Sb$_5$, as shown in Figs.~\ref{fig:fig_nosign}(a)-(b), a series of dI/dV spectra are taken across nonmagnetic and magnetic defects (green lines in the insets). The superconducting gap around nonmagnetic Zn adatoms and its vicinity remains unchanged, while for the magnetic Cr clusters, the superconducting gap is significantly suppressed and a pair of asymmetric peaks appear inside the gap, which are hallmarks of impurity induced in-gap states. Away from the Cr cluster, the impurity states weaken, and the superconducting gap gradually recovers(Figs.~\ref{fig:fig_nosign}(b)). These observations of SC being sensitive to magnetic impurities and robust against nonmagnetic impurities provide compelling evidence for sign-preserving in CsV$_3$Sb$_5$. 

The conclusion of nodeless and sign-preserving SC in CsV$_3$Sb$_5$ is also obtained in a study using electron irradiation to tune the concentration of nonmagnetic impurities \cite{roppongi2023bulk}. With increasing impurities from the pristine sample to the 8.6 C/cm$^2$ irradiated sample, measurements of $\Delta\lambda(T)$ suggest persistent fully-gapped behavior, indicating that nodeless SC is robust against nonmagnetic impurities (Figs.~\ref{fig:fig_nosign}(c)-(d)). This also rules out a fully-gapped sign-changing order parameter such as $s\pm$ in CsV$_3$Sb$_5$, in which case $\Delta\lambda(T)$ is expected to change to a $T^2$ dependence in the irradiated sample \cite{prozorov2011london}. 

Another experimental hallmark of sign-preserving SC is the Hebel-Slichter peak in NMR measurements of the spin-lattice relaxation rate. The observation of a Hebel-Slichter peak offers strong evidence for sign-preserving SC, although its absence does not necessarily indicate sign-changing SC. In the case of CsV$_3$Sb$_5$, a clear Hebel-Slichter peak is observed just below $T_{\rm c}$ (Fig.~\ref{fig:fig_nosign}(h)), evidencing a sign-preserving superconducting order parameter \cite{mu2021s}. 

The centrosymmetric structure of $A$V$_3$Sb$_5$ (Figs.~\ref{fig:fig1}(a)-(b)) rules out the possibility of a mixture of spin-singlet and spin-triplet components in the superconducting pairing state\cite{smidman2017superconductivity}. This consideration in combination with the reduction of the Knight shift in the spin-lattice relaxation rate of CsV$_5$Sb$_5$ below $T_{\rm c}$ for applied fields along various high symmetry directions \cite{mu2021s} (Fig.~\ref{fig:fig_nosign}(e)-(g)), offers compelling evidence of spin-singlet SC in $A$V$_3$Sb$_5$.

\begin{figure}[htbp]
    \centering
    \includegraphics[width=\columnwidth]{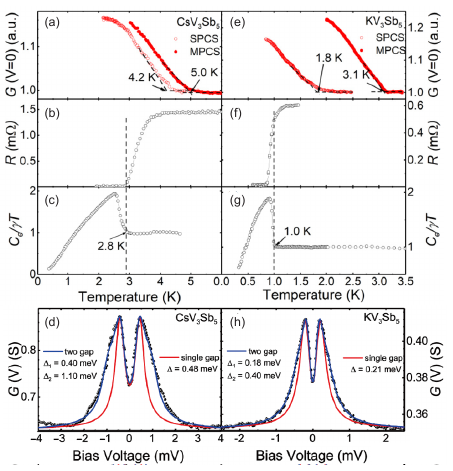}
    \caption{
      \label{PCS_zbc_and_didv}
      (a) The temperature dependence of zero-bias conductance measured by PCS on $\mathrm{CsV_3Sb_5}$, in comparison with measurements of (b) resistance and (c) heat capacity on the same crystal. Representative SPCS differential conductance curves of (d) $\mathrm{CsV_3Sb_5}$ measured at $T=0.3$~K, in comparison with fits to $s$-wave single-gap (red) and two-gap (blue) models. Similar measurements for KV$_3$Sb$_5$ are presented in (e)-(h)\cite{yin2021strain}.}
  \end{figure}

Point-contact spectroscopy (PCS) is another effective experimental technique for probing the superconducting order parameter \cite{le2022fermi}. Soft PCS (SPCS) generally uses conductive silver glue to make contact with the sample, while mechanical PCS (MPCS) adjusts the relative position between a sharp tip and the sample using a nano-manipulator, allowing for control of the contact area and strain on the sample. In SPCS and MPCS measurements on $\mathrm{CsV_3Sb_5}$ and $\mathrm{KV_3Sb_5}$, it was found that  $T_{\rm c}$ determined from specific heat (Figs.~\ref{PCS_zbc_and_didv}(c) and (g)) and electrical transport (Figs.~\ref{PCS_zbc_and_didv}(b) and (f)) are significantly lower than $T_{\rm c}$ determined from the onset of Andreev reflection in PCS measurements (Figs.~\ref{PCS_zbc_and_didv}(a) and (e)) \cite{yin2021strain}.  Moreover, the values of  $T_{\rm c}$ determined from MPCS are consistently higher than those from SPCS.

Given that such an enhancement of $T_{\rm c}$ is not observed by STM measurements \cite{liang2021three}, and that $\mathrm{CsV_3Sb_5}$ is a good metal with a large density of states near the Fermi level, the enhanced $T_{\rm c}$  seen by PCS measurements are unlikely due to surface states or carrier doping via the contact point. Instead, given the SC in $\mathrm{CsV_3Sb_5}$ is sensitive to pressure and strain \cite{chen2021double,yu2021unusual,qian2021revealing,yang2023plane} (see Section 3 for more details),  the enhanced $T_{\rm c}$ in PCS likely arises from by local stress or strain at the contact point. As the stress or strain in MPCS is generally larger than that in SPCS, this scenario also accounts for the larger $T_{\rm c}$ enhancement in MPCS.

The SPCS differential conductance curves of $\mathrm{CsV_3Sb_5}$ and $\mathrm{KV_3Sb_5}$ consist of a pair of symmetric peaks (Figs.~\ref{PCS_zbc_and_didv}(d) and (h)), which cannot be adequately described by a single-gap $s$-wave model (red lines), and instead require two $s$-wave gaps (blue lines) \cite{yin2021strain}. For $\mathrm{CsV_3Sb_5}$, the superconducting coupling strengths for the two gaps are $2\Delta_1/{k}_{\rm B}T_{\rm c} =$ 2.3 and $2\Delta_2/k_{\rm B}{T}_{\rm c}$ = 7.4 (using the enhanced values of $T_{\rm c}$ in Fig.~\ref{PCS_zbc_and_didv}(a)), significantly larger than values determined in TDO measurements \cite{duan2021nodeless}. This suggests that while nodeless $s$-wave SC is robust under stress or strain, the superconducting coupling strength can be sensitively tuned. The coupling strengths in $\mathrm{KV_3Sb_5}$  are found to be $2\Delta_1/{k}_{\rm B}T_{\rm c} =$ 2.3 and $2\Delta_2/k_{\rm B}{T}_{\rm c}$ = 5.2 \cite{yin2021strain}. 

In contrast to SPCS measurements, the differential conductance curves in MPCS are well described by a single-gap $s$-wave model, with $2\Delta/{k}_{\rm B}{T}_{\rm c} = 2.2$ for $\mathrm{CsV_3Sb_5}$ and $2\Delta/{k}_{\rm B}{T}_{\rm c} = 2.5$ for $\mathrm{KV_3Sb_5}$ \cite{yin2021strain}. These values are close to those of the smaller gap in SPCS, and suggests that the larger gap is not detected in MCPS. Since PCS with current along the $c$-axis cannot detected two-dimensional superconducting gaps, a distinct possibility is that the larger strain in MPCS tunes the Fermi surface associated with the larger gap to become two-dimensional. These findings in PCS measurements suggest strain not only tunes $T_{\rm c}$, but also strongly affects the coupling strengths and Fermi surfaces in the kagome metals $A$V$_3$Sb$_5$.

Table \ref{Table_gap} summarizes the pairing symmetry and superconducting coupling strengths of $\mathrm{CsV_3Sb_5}$ studied using various techniques. Overall, these studies consistently find $\mathrm{CsV_3Sb_5}$ to be a nodeless multi-gap superconductor, although the obtained superconducting coupling strengths vary between studies. For instance, TDO detects a relatively small superconducting coupling strength, while PCS experiments report coupling strengths that are significantly higher than the weak-coupling limit\cite{duan2021nodeless,yin2021strain,he2022strong}. The coupling strengths obtained in different STM measurements also vary \cite{xu2021multiband,liang2021three}, which may result from SC gaps associated with different bands being detected, or inadvertent strain which tunes the coupling strength, varying between measured samples. 
It is interesting to note that STM measurements uncover both V-shaped and U-shaped gaps, with the V-shaped gaps resulting from two $s$-wave gaps and the U-shaped gaps resulting from a single $s$-wave gap \cite{xu2021multiband}. 

\begin{figure}[h]
  \centering
  \includegraphics[width=\columnwidth]{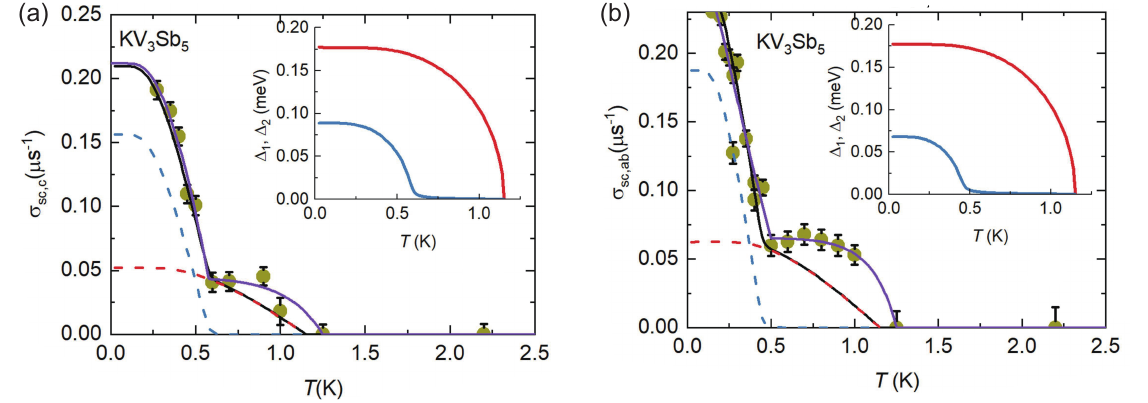}
  \caption{
  Temperature dependence of the superconducting contribution to muon relaxation rates (a) $\sigma_{SC,c}$ and (b) $\sigma_{SC,ab}$ in KV$_3$Sb$_5$measured under 5~mT fields perpendicular and parallel to the $ab$-plane, respectively \cite{mielke2022time}.
  }
    \label{fig5_R_K135}
\end{figure}

Compared to CsV$_3$Sb$_5$, there is yet no consensus on the SC gap structure of RbV$_3$Sb$_5$ and KV$_3$Sb$_5$. $\mu$SR measurements of the superconducting contribution to muon relaxation rates $\sigma_{SC,ab}$ and $\sigma_{SC,c}$ (which are related with the magnetic penetration depth through $\sigma_{SC}(T)\propto\lambda^{-2}(T)$) in KV$_3$Sb$_5$ find pronounced two-step behaviours \cite{mielke2022time}. Such a two-step behavior is modelled by the $\gamma$ model, where two bands each with $s$-wave gaps have an extremely weak interband coupling and independent values of $T_{\rm c}$ (Fig.\ref{fig5_R_K135} (a)(b)). This is qualitatively consistent with PCS measurements on KV$_3$Sb$_5$ that find two $s$-wave gaps at low temperatures \cite{yin2021strain}. On the other hand, a different $\mu$SR studies finds both RbV$_3$Sb$_5$ and KV$_3$Sb$_5$ exhibit nodal SC at ambient pressure, which evolve to become nodeless under pressure \cite{guguchia2023tunable}. In comparison with the numerous studies of pairing symmetry in CsV$_3$Sb$_5$, the limited number of similar studies on RbV$_3$Sb$_5$ and KV$_3$Sb$_5$ motivate further examinations of the pairing symmetries in these $A$V$_3$Sb$_5$ variants.

\begin{table*}[h]
\begin{center}
   \caption{\label{Table_gap}Superconducting gaps and coupling strengths of CsV$_3$Sb$_5$ probed by various experimental techniques\cite{zhang2023superconducting}.}
\scalebox{0.72}{
\begin{tabular}{c|c|c|c}
\midrule
 \textbf{Technique}     &  \textbf{Model}                  &  \textbf{2$\Delta_1/k_BT_c$}                                                                                                                                        &  \textbf{2$\Delta_2/k_BT_c$}                                                                  \\[0.5ex] \hline
Heat capacity \cite{duan2021nodeless}& two $s$-wave& 1.26                                                                                                                                          & 3.24                                                                     \\ \hline
TDO \cite{duan2021nodeless}& two $s$-wave& 1.16                                                                                                                                          & 2.84                                                                     \\ \hline
TDO \cite{roppongi2023bulk}& \begin{tabular}[c]{@{}c@{}}anisotropic s\\ \\ + isotropic s-wave\end{tabular} & \begin{tabular}[c]{@{}r@{}}0.88($\Delta_{min}$)\\ \\ 4.1($\Delta_{max}$)\end{tabular}                                                                                     & 2.48                                                                     \\ \hline
$\mu$SR \cite{shan2022muon}& two $s$-wave& 1.1                                                                                                                                           & 5.44                                                                     \\ \hline
$\mu$SR \cite{gupta2022microscopic}& two $s$-wave& 2.22                                                                                                                                          & 5.5                                                                      \\ \hline
$\mu$SR \cite{gupta2022two}& two $s$-wave& 2.78                                                                                                                                          & 3.71                                                                     \\ \hline
STM \cite{xu2021multiband}& multiple $s$-wave& \multicolumn{1}{c|}{\begin{tabular}[c]{@{}c@{}} (Sb-surface)\\ V-gap 3.2\\ U-gap 3.4\\ (Half-Cs-surface)\\ V-gap 2.7\\ U-gap 4.0\end{tabular}} & \begin{tabular}[c]{@{}c@{}} \\4.3\\/\\ \\5.1\\/\end{tabular}             \\ \hline
STM \cite{liang2021three}& \begin{tabular}[c]{@{}c@{}} \\one anisotropic s\\ \\or two s-wave\end{tabular} & \multicolumn{1}{c|}{\begin{tabular}[c]{@{}c@{}} 1.19($\Delta_{min}$),\\  \\2.97($\Delta_{max}$) \\2.17\end{tabular}}                                                        & \multicolumn{1}{c}{\begin{tabular}[c]{@{}l@{}} \\/\\ \\ 2.61 \end{tabular}} \\
\hline
SPCS \cite{he2022strong}& two $s$-wave& \multicolumn{1}{c|}{3.1}                                                                                                                      & 7.2                                                                      \\ \hline
SPCS \cite{yin2021strain}& two $s$-wave& \multicolumn{1}{c|}{2.3}                                                                                                                      & 7.4                                                                      \\ \hline
MPCS \cite{yin2021strain}& single $s$-wave& \multicolumn{1}{c|}{2.2}                                                                                                                      & /                                                                        \\ \hline
\begin{tabular}[c]{@{}c@{}} self-field  critical current (thin flakes) \cite{zhang2023nodeless}\end{tabular}& single $s$-wave& \multicolumn{1}{c|}{5.4-5.68}                                                                                                                 & /                                                                        \\ \midrule
\end{tabular} }
\end{center}
\end{table*}

\section{Evolution of CDW and SC under pressure}
Hydrostatic pressure plays a crucial role in the research of SC, strongly correlated electronic systems, and topological states of matter. In heavy fermion systems such as CeCu$_2$Si$_2$ and CeRhIn$_5$, pressure can be used to access superconducting, quantum critical, and non-Fermi liquid phases \cite{hegger2000pressure,park2006hidden,yuan2003observation,weng2016multiple}. For instance, a ferromagnetic quantum critical point was recently uncovered in the heavy fermion system CeRh$_6$Ge$_4$, challenging the long-held belief that such phenomenon could not be achieved in systems with minimal disorder 
\cite{shen2020strange}. In systems where CDW and SC coexist, pressure has been widely used to examine the interplay between the two orders and how SC evolves upon tuning \cite{gruner2017charge,goh2015ambient,gabovich2001charge,du2020interplay,shen2020evolution,klintberg2012pressure}, motivating similar studies of $A$V$_3$Sb$_5$ under pressure.

\subsection{The interplay of CDW and SC at low pressures}

\begin{figure}[htbp]
	\includegraphics[width=\linewidth]{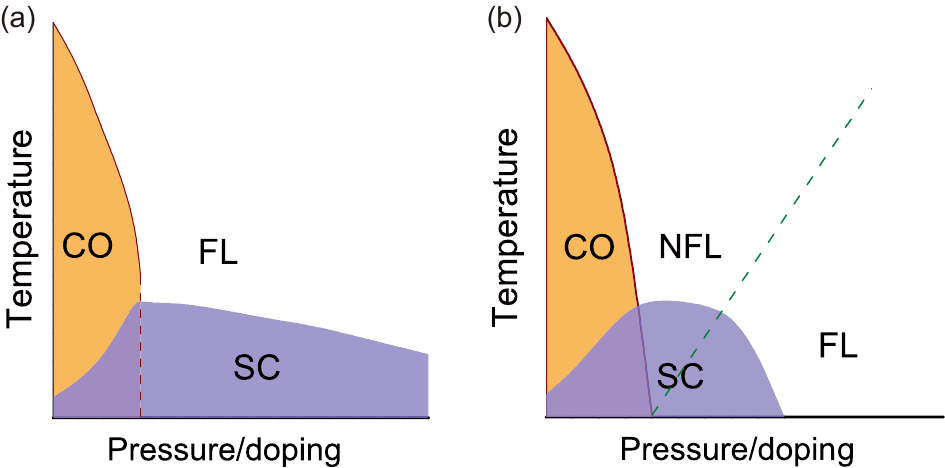}
	\caption{Schematic phase diagrams of the interplay between CDW and SC \cite{du2021pressure}. (a) Competition between CDW and SC, and the system can be adequately described as a Fermi liquid (FL). (b) Quantum fluctuations near a CDW quantum	critical point promotes superconducting pairing, and leads to a regime of non-Fermi-lquid (NFL) behavior, similar to SC arising near magnetic quantum critical points.}
	\label{scenario}
\end{figure}

There are two main possible scenarios for the interplay between SC and CDW: (1) SC competes with CDW, and (2) SC is enhanced by quantum critical CDW fluctuations. In the first scenario (Fig.~\ref{scenario}(a)), CDW leads to gapping of the Fermi surface and a reduced electronic density of states, inhibiting superconductivity \cite{gabovich2001charge,du2020interplay,shen2020evolution}. Due to their competing nature, when CDW is suppressed through the application of pressure or doping, $T_{\rm c}$ significantly increases. After the full suppression of CDW, $T_{\rm c}$ evolves more gradually upon tuning. In the second scenario (Fig.~\ref{scenario}(b)), quantum critical fluctuations near a CDW quantum critical point promotes superconducting pairing. This results in a superconducting dome with maximum $T_{\rm c}$ around the quantum critical point, similar to SC domes observed around magnetic quantum critical points in heavy fermions and iron-based superconductors \cite{yuan2003observation,mathur1998magnetically,dai2009iron,shibauchi2014a}. In material systems hosting coexistent CDW and SC studied thus far, in most cases the two orders compete (scenario 1) , whereas reports of systems that fall into scenario 2 have been limited \cite{gruner2017charge,goh2015ambient,klintberg2012pressure}.

\begin{figure*}[htbp]
	\centering
	\includegraphics[width=0.8\linewidth]{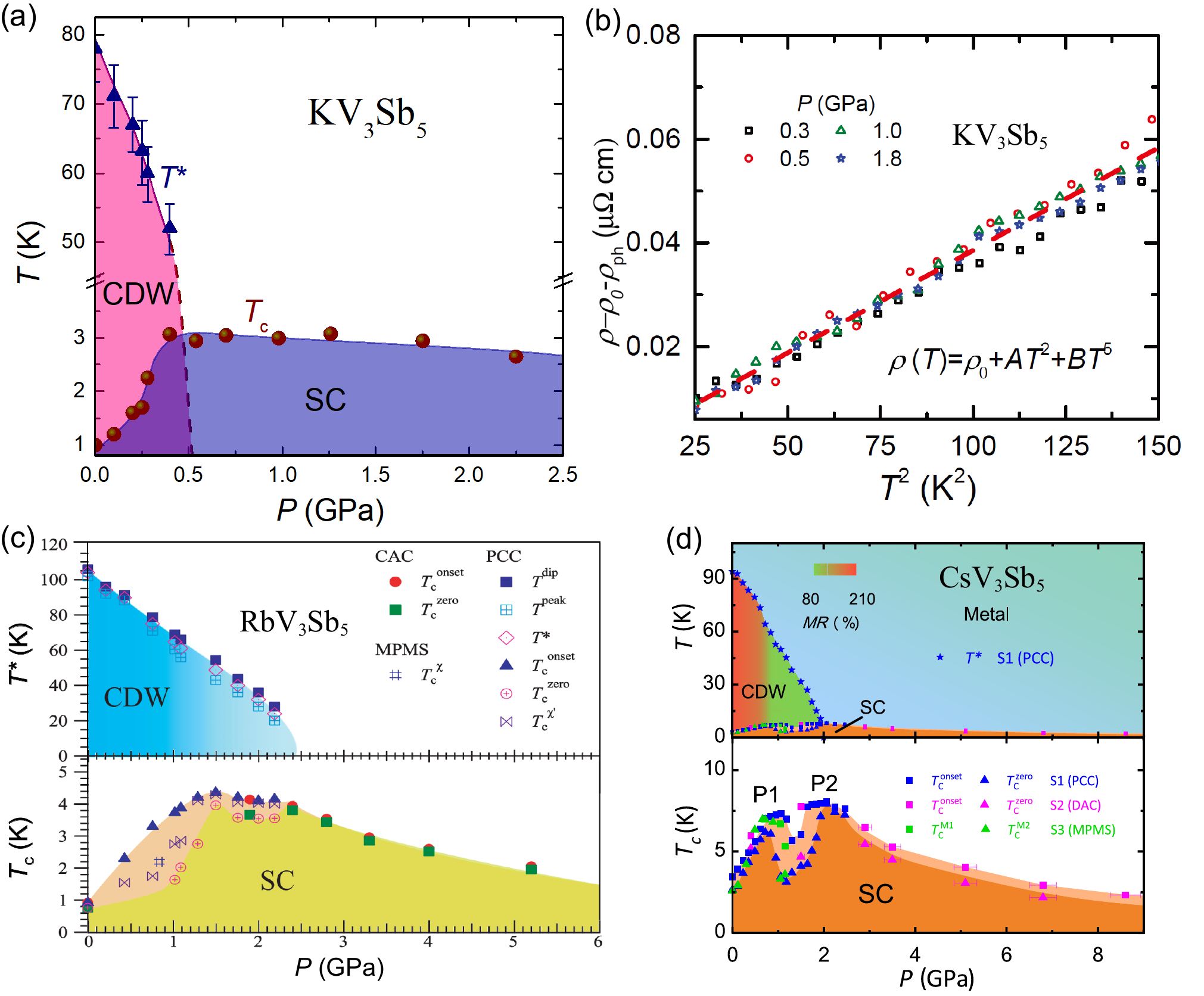}
	\caption{(a) $P$-$T$ phase diagram of KV$_3$Sb$_5$ below 2.5~GPa \cite{du2021pressure}. (b) Resistivity of KV$_3$Sb$_5$ plotted as a function of $T^2$, after subtracting the residual resistivity $\rho_0$ and phonon contributions \cite{du2021pressure}. (c) $P$-$T$ phase diagram of RbV$_3$Sb$_5$ below 6~GPa \cite{wang2021competition}. (d) $P$-$T$ phase diagram of CsV$_3$Sb$_5$ below 9~GPa \cite{yu2021unusual}.}
	\label{lowpPD}
\end{figure*}

The interplay between CDW and SC in $A$V$_3$Sb$_5$ has been extensively studied \cite{du2021pressure,du2022evolution,zhang2021first,song2023anomalous,chen2021double,yu2021unusual,hou2023effect,qian2021revealing,kang2023charge,consiglio2022van,li2022discovery,guguchia2023tunable,oey2022tuning,song2021competition,yang2022titanium,lin2022multidome,li2022tuning,si2022charge,wang2022charge,wang2021competition,zheng2022emergent}. Through pressure tuning, these studies consistently reveal a competition between CDW and SC in $A$V$_3$Sb$_5$. Taking KV$_3$Sb$_5$ as an example \cite{du2021pressure}, $T_{\rm CDW}$ gradually decreases with increasing pressure (Fig.~\ref{lowpPD}(a)). At around $P_{\rm c}$ $\sim0.5$~GPa, the CDW abruptly disappears through a first-order quantum phase transition, without signatures of a CDW quantum critical point. Concurrently, the superconducting transition temperature $T_{\rm c}$ increases for pressures up to  $P_{\rm c}$, where a maximum of 3.1 K is reached, $T_{\rm c}$ then slowly decreases with further increase of pressure. By subtracting the residual resistivity and phonon scattering contributions from the normal state resistivity of KV$_3$Sb$_5$ [Fig.~\ref{lowpPD}(b)], it was found that the system exhibits typical Fermi liquid behavior between 0.3~GPa and 1.8~GPa, and the coefficients of the $T^2$ term are similar at different pressures. These observation suggest that the electron-electron scattering rate does not have a significant pressure dependence, and further supports the absence of a CDW quantum critical point in pressurized KV$_3$Sb$_5$. A similar competition between CDW and SC is found in RbV$_3$Sb$_5$ \cite{du2022evolution}.

The interplay between CDW and SC in CsV$_3$Sb$_5$ is more complex \cite{chen2021double,yu2021unusual}. $T_{\rm CDW}$ is monotonically suppressed with increasing pressure until it disappears in a first-order fashion at $P_{\rm 2}\sim2.0$~GPa. At the same time, $T_{\rm c}$ increases from its ambient pressure value of 2.7~K to 8~K at $P_{\rm 2}$. However, within the pressure range where superconductivity and CDW coexist, $T_{\rm c}$ shows another maximum at $P_{\rm 1}\sim0.7$~GPa and exhibits a local minimum between $P_{\rm 1}$ and $P_{\rm 2}$. This results in an unusual double dome SC in the $P-T$ phase diagram of CsV$_3$Sb$_5$ at low pressures ($P\leq$3 GPa) (Fig.~\ref*{lowpPD}(d)). It is worth noting that a similar two-dome SC is suggested for RbV$_3$Sb$_5$ at low pressures (Fig.~\ref*{lowpPD}(c)), with the CDW disappearing around 2.4~GPa and two $T_{\rm c}$ maxima at $P_{\rm 1}\sim1.5$~GPa and $P_{\rm 2}\sim2.4$~GPa \cite{wang2021competition}. Compared to CsV$_3$Sb$_5$, the two $T_{\rm c}$ maxima in RbV$_3$Sb$_5$ are much less prominent, an no such two $T_{\rm c}$ maxima are seen in KV$_3$Sb$_5$ at low pressures \cite{du2021pressure}.

\begin{figure*}[htbp]
  \centering
  \includegraphics[width=0.8\textwidth]{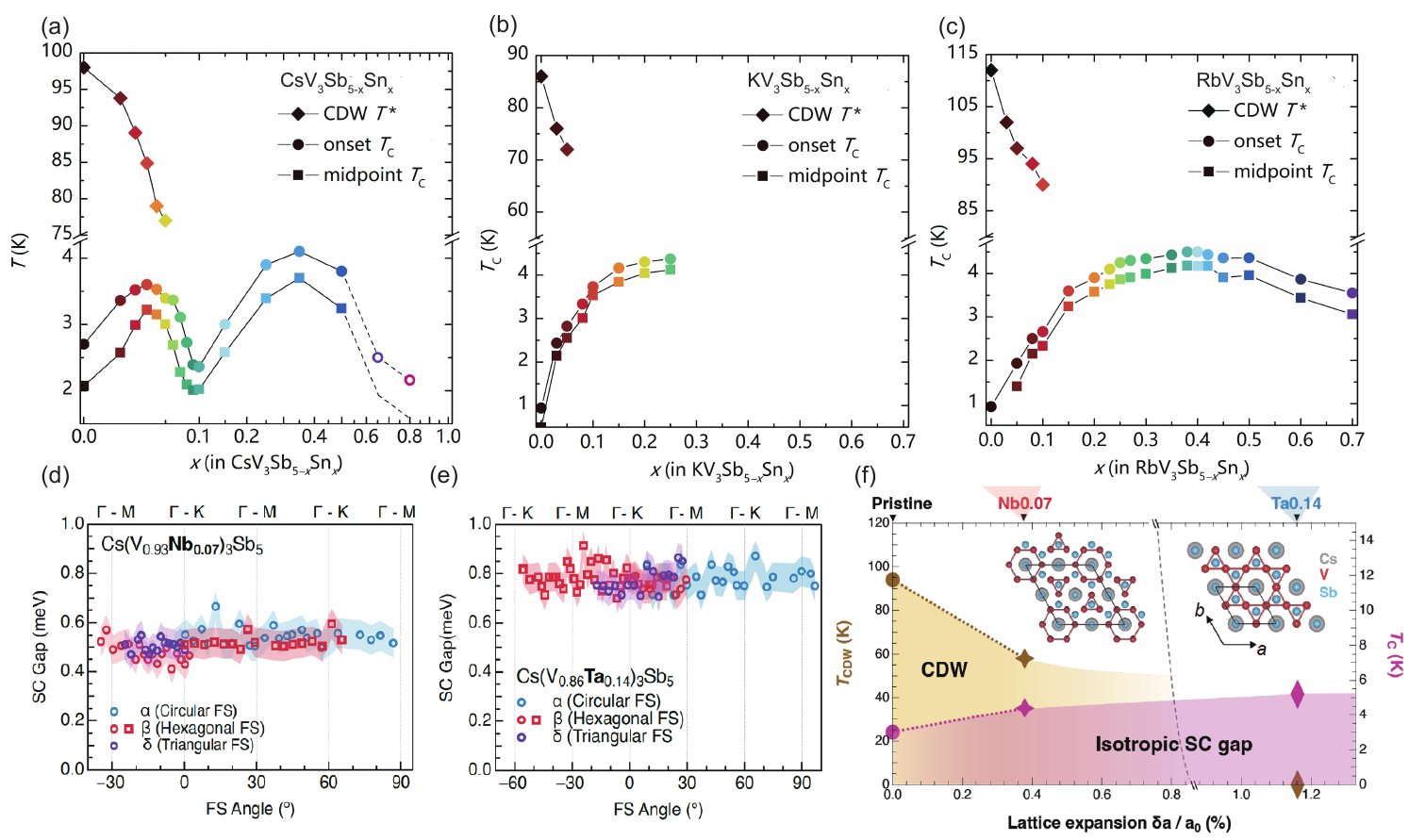}
  \caption{
  (a)-(c) The Sn-substitution phase diagrams of $A$V$_3$Sb$_5$ ($A$ = Cs, K, Rb) \cite{oey2022fermi,oey2022tuning}. 
   Superconducting gaps for various Fermi surfaces and directions, measured using ARPES for (d) $\mathrm{Cs(V_{0.93}Nb_{0.07})_3Sb_5}$ and (e) $\mathrm{Cs(V_{0.86}Ta_{0.14})_3Sb_5}$. (f) Evolution of CDW and SC in CsV$_3$Sb$_5$, as a function of lattice expansion introduced by isovalent dopants. Regardless of whether the CDW is present, isovalent-doped $\mathrm{CsV_3Sb_5}$ samples exhibit isotropic nodeless superconducting gaps \cite{zhong2023nodeless}.
  }
    \label{fig:fig_doping2}
\end{figure*}

Compared with hydrostatic pressure, chemical substitution not only introduces chemical pressure, but also carrier doping and disorder effects. By introducing holes or electrons into $A$V$_3$Sb$_5$, band filling is expected to be tuned such that a vHs can move towards or away from the Fermi level, which in turn affects the CDW and offers another means to probe the interplay between CDW and SC. One way to achieve hole-doping is through the substitution of Sn for Sb, the resulting phase diagram of CsV$_3$Sb$_{5-x}$Sn$_x$ exhibits two SC domes: one that coexists with CDW and another that emerges when the long-range CDW is fully suppressed, replaced by a short-range CDW(Fig.~\ref{fig:fig_doping2}(a))\cite{oey2022fermi,kautzsch2023incommensurate}. In contrast to CsV$_3$Sb$_{5-x}$Sn$_x$, a single dome is observed in KV$_3$Sb$_5$ and RbV$_3$Sb$_5$ upon Sn substitution (Figs.~\ref{fig:fig_doping2}(b)-(c)) \cite{oey2022tuning}. Such a difference is reminiscent of the behavior under hydrostatic pressure, and suggests the emergence of new CDW states that compete more strongly with SC in pressurized and doped CsV$_3$Sb$_5$. For the pressure-induced CDW in CsV$_3$Sb$_5$, whereas NMR measurements find evidence for a stripe-like or an incommensurate CDW \cite{zheng2022emergent,feng2023commensurate}, XRD measurements reveal it to be a long-range commensurate CDW modulated by the wave vector $(0,3/8,1/2)$ \cite{stier2024pressure}. For Sn-doped CsV$_3$Sb$_5$, XRD measurements reveal that with increasing Sn content, the long-range CDW in CsV$_3$Sb$_5$ gives way to a stripe-like short-range CDW\cite{kautzsch2023incommensurate}. One scenario that may unify these observations is that while the $(0,3/8,1/2)$ CDW in pressurized CsV$_3$Sb$_5$ is long-range, it has stripe-like low-lying excited states, which can become pinned by disorder introduced by Sn-doping. Phonon measurements in the $(0,3/8,1/2)$ CDW of pressurized CsV$_3$Sb$_5$ and the stripe-like short-range CDW in Sn-doped CsV$_3$Sb$_5$ will be helpful to clarify their similarities and differences.

Hole-doping can also be achieved through the substitution of V in the kagome layer with Ti, which leads to the suppression of CDW as in CsV$_3$Sb$_{5-x}$Sn$_x$\cite{sur2023optimized,liu2023doping,yang2022titanium}. Also similar to Sn-doped CsV$_3$Sb$_5$, two SC domes separated by a $T_{\rm c}$ minimum (around $x\approx0.02$) are observed in Cs(V$_{1-x}$Ti$_x$)$_3$Sb$_5$ \cite{sur2023optimized,yang2022titanium}. From a rigid band perspective, as hole-doping moves the vHs closer to the Fermi level and the nesting between vHss promotes the CDW state, $T_{\rm CDW}$ is expected to increase upon hole-doping, which is contrary to experimental observations. This in turn implicates a role of disorder effects that weaken the CDW in hole-doped CsV$_3$Sb$_5$. Such disorder effects likely also play a role in the evolution from long-range towards short-range CDW with increasing hole-doping \cite{kautzsch2023incommensurate}.

The substitution of V with Ta, Nb, which do not lead to carrier doping, has also been explored in CsV$_3$Sb$_5$. The phase diagram in these cases only exhibits a single SC dome \cite{liu2022evolution,li2022tuning,zhong2023nodeless} and shows orbital-independent, nearly isotropic SC gap in momentum space (Fig.~\ref{fig:fig_doping2}(f)). Compared to 14\% Ta-substituted CsV$_3$Sb$_5$, the 7\% Nb-subsituted sample exhibits a smaller superconducting gap, which results from remnant CDW in the latter that weakens SC(Fig.~\ref{fig:fig_doping2}(d)-(e)). In the case of electron-doping via substituting V with Cr \cite{ding2022effect} and Mo\cite{liu2022evolution}, SC is rapidly suppressed in both cases, while CDW is suppressed slowly by Cr and enhanced by Mo, suggesting effects beyond carrier doping and chemical pressure in tuning CDW and SC. These results underscore the importance of disorder effects and impurity potential strengths in determining the physical properties of doped $A$V$_3$Sb$_5$ systems.    

\begin{figure}
    \centering
    \includegraphics[width=\columnwidth]{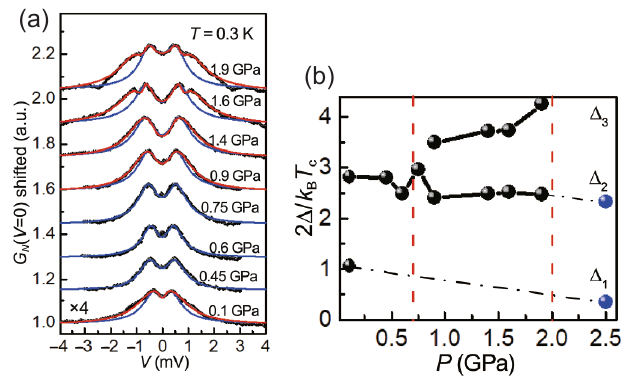}
    \caption{
      \label{PCS_pressure}
      The pressure evolution of superconducting properties in $\mathrm{CsV_3Sb_5}$ probed by SPCS \cite{zhang2023superconducting}. (a) The normalized differential conductance curves at 0.3 K and various pressures measured by SPCS, compared with fits to $s$-wave single-gap (blue) and two-gap (red) models.
      (b) Superconducting coupling strengths in $\mathrm{CsV_3Sb_5}$ as a function of pressure.
    }
  \end{figure}

\begin{figure}
    \centering
    \includegraphics[width=\columnwidth]{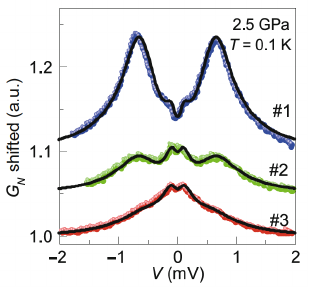}
    \caption{
      \label{PCS_2p5}
      Representative normalized SPCS differential conductance curves obtained for various point contacts on the same $\mathrm{CsV_3Sb_5}$ sample, measured at 2.5 GPa and 0.1 K \cite{zhang2023superconducting}. Black lines represent fits to a two-gap $s$-wave model.
      }
      \end{figure}

Given the unusual evolution of SC and the emergence of a new CDW phase in pressurized CsV$_3$Sb$_5$, it is important to elucidate how the superconducting gap and coupling strength evolve under pressure. To address these questions, spectroscopic techniques that are compatible with hydrostatic pressure are needed. In this regard, the simplicity and versatility of SPCS makes it an ideal technique to probe the evolution of SC in CsV$_3$Sb$_5$ under pressure \cite{zhang2023superconducting}. Differential conductance curves at 0.3 K reveal a systematic and nontrivial evolution of superconducting gaps under pressure (Fig. \ref*{PCS_pressure} (a)). At 0.1 GPa, the differential conductance curve reveals a two-gap $s$-wave behavior (solid red line), which becomes a single-gap $s$-wave behavior for pressures ranging from 0.1 GPa to $P_1$. When the pressure is further increased to $P_1 < P < P_2$, the differential conductance curves exhibit an additional broad feature at higher biases, implying the emergence of a larger gap which becomes more prominent with increasing pressure up to 1.6 GPa (Fig. \ref*{PCS_pressure} (a)). At 2.5 GPa ($P > P_2$), the conductance curves become significantly modified compared to 1.9~GPa, with two clear peaks at 0.1~mV and~0.7 mV (Fig. \ref*{PCS_2p5}) that can be described by two $s$-wave gaps. The SPCS measurements reveal that $\mathrm{CsV_3Sb_5}$ displays $s$-wave superconductivity regardless of whether there is a coexistent CDW, although the probed gaps vary strongly in size between different pressure regimes. 

The superconducting coupling strengths at different pressures are summarized in Fig. \ref*{PCS_pressure} (b), and three gaps with $2\Delta_1/{k}_{\rm B}{T}_{\rm c} \lesssim 1$, $2\Delta_2/{k}_{\rm B}{T}_{\rm c} \approx 2.3 - 3.0$, $2\Delta_3/{k}_{\rm B}{T}_{\rm c} \gtrsim 3.5$ can be identified. The smallest gap $\Delta_1$ is only clearly detected at 0.1~GPa and 2.5~GPa (Fig. \ref*{PCS_2p5}), whereas for pressures in between, $\Delta_1$ likely becomes significantly smaller and leads to weak conductance peaks near zero bias (0.6~GPa and 0.75~GPa in Fig. \ref*{PCS_pressure}(a)). $\Delta_2$ is identified throughout the entire pressure range below 2.5~GPa, and this gap is consistently detected in ambient pressure CsV$_3$Sb$_5$ and its doped variants through a variety of experimental techniques \cite{liang2021three,shan2022muon,duan2021nodeless,xu2021multiband,gupta2022microscopic,gupta2022two,roppongi2023bulk,zhong2023nodeless}. The largest gap $\Delta_3$ is exclusively observed in the pressure range $P_1 < P < P_2$, although it should be present across the entire pressure range considering that the superconducting coupling strengths of $\Delta_1$ and $\Delta_2$ are both smaller than the weak-coupling limit of 3.52. The absence of $\Delta_3$ for $P < P_1$ and $P > P_2$ is likely a consequence of the two-dimensional character of its corresponding Fermi surface at these pressures, which cannot be probed by a point contact with current along the $c$-axis.

In the pressure range $P_1 < P < P_2$, ${T}_{\rm c}$ is suppressed, the superconducting transition is significantly broadened, the residual resistivity is enhanced, the magnetoresistance is reduced, and quantum oscillations rapidly decays \cite{yu2021unusual}, pointing to a unique interplay between SC and CDW. NMR and $\mu$SR measurements\cite{zheng2022emergent,gupta2022two} evidence a new CDW, possibly consisting of stripes, emerge within this pressure range, competes more strongly with the SC and leads to a decrease in ${T}_{\rm c}$. The SPCS measurements further indicate this new CDW state modifies the two-dimensional Fermi surface associated with $\Delta_3$ to become three-dimensional in the pressure range $P_1 < P < P_2$. Increasing pressure above $P_2$ which completely suppresses the CDW, the Fermi surface associated with $\Delta_3$ changes from three-dimensional to two-dimensional, rendering it once again undetectable by SPCS. 

The persistence of $s$-wave superconductivity in $\mathrm{CsV_3Sb_5}$ under hydrostatic pressure is also reported in other SPCS studies \cite{he2022strong,wen2023emergent}. Compared to Ref.~\cite{zhang2023superconducting} that finds multi-gap behavior, the differential conductance curves in Ref.~\cite{wen2023emergent} are adequately described by a single-gap model. This discrepancy may result from a slighter higher measurement temperature in Ref.~\cite{wen2023emergent}, which could make it difficult to resolve different superconducting gaps.
The presence of multiple $s$-wave gaps in  $\mathrm{CsV_3Sb_5}$ under pressure is also observed in $\mu$SR measurements \cite{gupta2022two}. 

In $\mu$SR studies of the evolution of SC gaps in RbV$_3$Sb$_5$ and  KV$_3$Sb$_5$ under pressure \cite{guguchia2023tunable} , it is proposed that the SC gaps become nodeless after the full suppression of CDW in $\mathrm{KV_3Sb_5}$ ($P\gtrsim$0.5~GPa) and a partial suppression of CDW in $\mathrm{RbV_3Sb_5}$ ($P\gtrsim1.5$~GPa), similar to the behavior of SC gaps in pressurized CsV$_3$Sb$_5$. This suggests that nodeless SC is common to $A$V$_3$Sb$_5$ without CDW. On the other hand, the SC gaps in $A$V$_3$Sb$_5$ with CDW appears to be material-dependent, with ambient pressure RbV$_3$Sb$_5$ and  KV$_3$Sb$_5$ showing nodal SC gaps \cite{guguchia2023tunable}, the origin of which remains to be understood. 

\subsection{Pressure-tuning of SC beyond the suppression of CDW}

Under high pressures accessible in a diamond anvil cell, which range from tens to hundreds of GPa, entirely new quantum states or physical phenomena may emerge. Examples include metallic hydrogen predicted by Ashcroft \cite{ashcroft1968metallic}, and SC with $T_{\rm c}$ near room temperature in hydrogen-rich compounds \cite{drozdov2019superconductivity,somayazulu2019evidence}.

For the $A$V$_3$Sb$_5$ compounds, a few GPa of pressure is sufficient to completely suppress the CDW, which is accompanied by a significant enhancement of SC. How SC evolve under higher pressures, and whether there are new physical mechanisms involved, are questions that have drawn interest in the research of $A$V$_3$Sb$_5$ .

\begin{figure*}[htbp]
	\includegraphics[width=\linewidth]{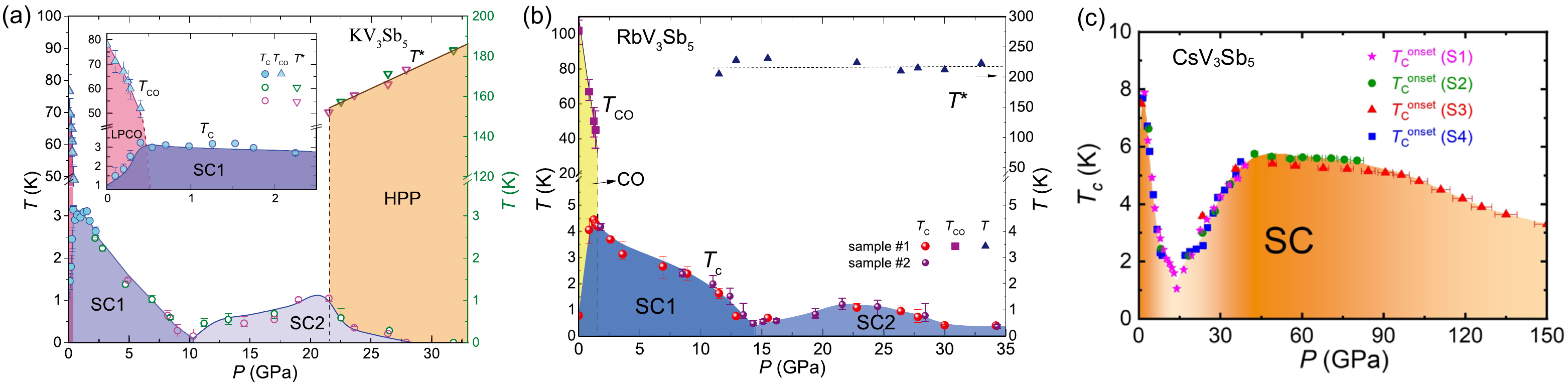}
	\caption{(a) High-pressure phase diagram of KV$_3$Sb$_5$ \cite{du2021pressure}. LPCO represents the low-pressure CDW, HPP represents a high-pressure phase, SC1 and SC2 represent two superconducting domes. (b) High-pressure phase diagram of RbV$_3$Sb$_5$ \cite{du2022evolution}. (c)High-pressure phase diagram of CsV$_3$Sb$_5$ \cite{yu2022pressure}.}
	\label{highpPD}
\end{figure*}

\begin{figure*}[htbp]
	\includegraphics[width=\linewidth]{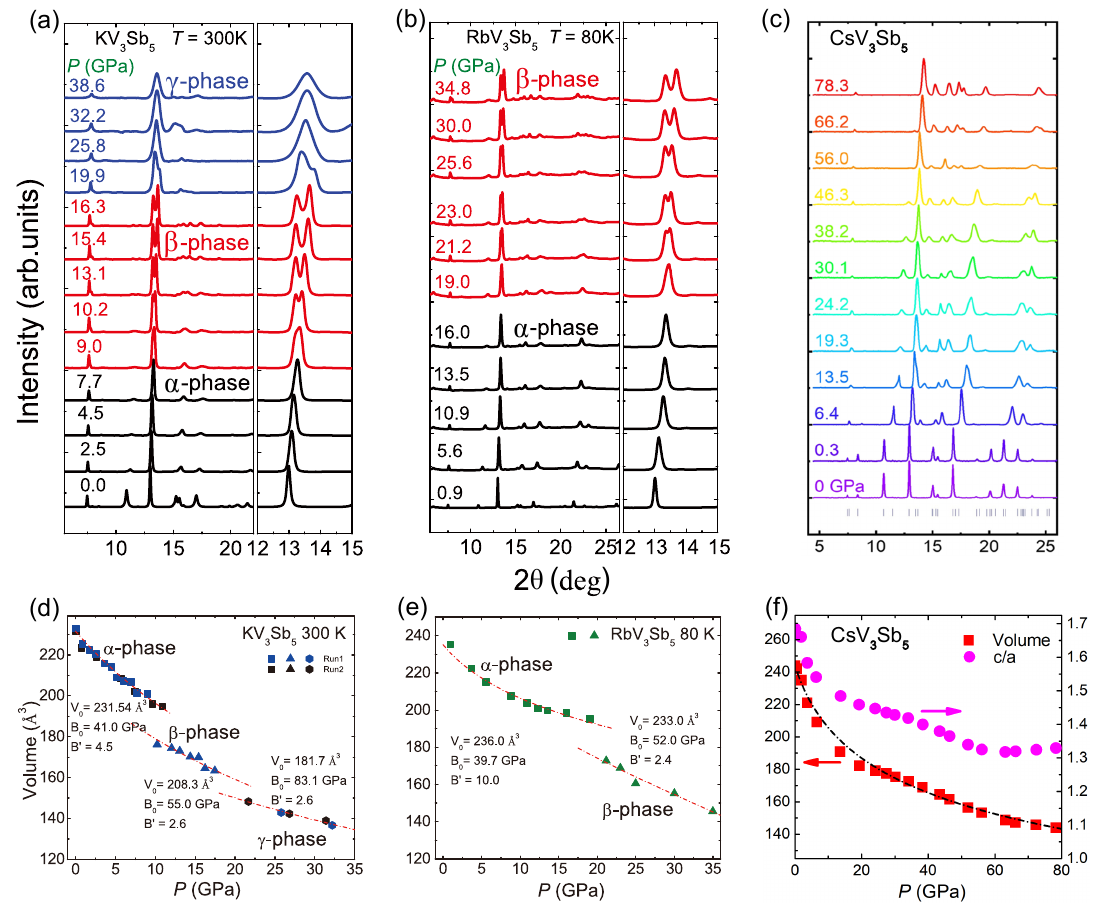}
	\caption{Pressure evolution of XRD data for (a) KV$_3$Sb$_5$, (b) RbV$_3$Sb$_5$, and (c) CsV$_3$Sb$_5$ \cite{du2022superconductivity,yu2022pressure}. The corresponding evolution of the unit cell volume of (d) KV$_3$Sb$_5$, (e) RbV$_3$Sb$_5$, and (f) CsV$_3$Sb$_5$. The ratio between lattice parameters $c$ and $a$ is also plotted in (f).}
	\label{XRD}
\end{figure*}

In electrical transport measurements of KV$_3$Sb$_5$ and RbV$_3$Sb$_5$ at high pressures, it is found that both compounds exhibit another superconducting dome beyond the full suppression of CDW (Figs.~\ref{highpPD}(a) and (b)). Taking KV$_3$Sb$_5$ as an example \cite{du2021pressure}, as pressure increases, $T_{\rm c}$ first rises and then falls (for pressures below 10~GPa), forming the first superconducting dome with a maximum $T_{\rm c}=3.1$~K at the pressure where CDW disappears. This indicates that the first superconducting dome results from the competition between SC and CDW. Beyond 10 GPa,  $T_{\rm c}$ rises again with increasing pressure, reaching a maximum of around 1 K at 22 GPa. With further increase of pressure,  $T_{\rm c}$ starts to decrease and SC disappears at around 28 GPa, forming the second superconducting dome. The SC properties of two domes are notably different, with the upper critical field being much higher in the second dome than that in the first dome, given a comparable $T_{\rm c}$ \cite{du2021pressure}. 

Furthermore, at pressures corresponding to the second superconducting dome, an anomaly in resistivity with clear thermal hysterisis is observed at high temperatures (denoted by $T^*$ in Fig.~\ref{highpPD}(a)). As this anomaly does not respond to a magnetic field, it is attributed to a first-order structural transition or a CDW transition. Interestingly, the difference between room temperature resistivity and residual resistivity tracks the evolution of $T_{\rm c}$ with pressure \cite{du2021pressure}, indicating an important role of electron-phonon interactions in determining the pressure evolution of SC in KV$_3$Sb$_5$.

Similar to KV$_3$Sb$_5$, additional superconducting domes beyond the full suppression of CDW are observed in RbV$_3$Sb$_5$ (Fig.~\ref{highpPD}(b)) \cite{du2022evolution} and CsV$_3$Sb$_5$ (Fig.~\ref*{highpPD}(c)) \cite{zhao2021nodal,zhang2021pressure,chen2021highly,zhu2022double,yu2022pressure}, suggestive of a common origin.

To elucidate the origin of the superconducting dome beyond the full suppression of CDW, high-pressure XRD measurements were conducted for KV$_3$Sb$_5$ and RbV$_3$Sb$_5$ \cite{du2022superconductivity}. As shown in Figs.~\ref{XRD}(a) and (d), KV$_3$Sb$_5$ maintains a hexagonal structure ($\alpha$ phase) up to 9~GPa, which separates the first and second superconducting domes. Above 9 GPa, the diffraction peak around 13° splits, indicating a pressure-induced structural phase transition ($\alpha$ phase to $\beta$ phase). Further increasing the pressure to 19.9 GPa, the split diffraction peaks at around 13° merge into a single peak again, signifying another pressure-induced structural phase transition ($\beta$ phase to $\gamma$ phase). Using the Le Bail method, it is found that the $\alpha$, $\beta$, and $\gamma$ phases respectively correspond to hexagonal, monoclinic, and orthorhombic structures. High-pressure XRD measurements for RbV$_3$Sb$_5$ also indicate a hexagonal ($\alpha$ phase) to monoclinic ($\beta$ phase) structural transition upon crossing into the second superconducting dome (Fig.~\ref{XRD}(b) and (e)). In contrast to KV$_3$Sb$_5$, RbV$_3$Sb$_5$ maintains the $\beta$ phase up to at least 35 GPa, without undergoing a structural phase transition into the $\gamma$ phase.

These experimental findings in KV$_3$Sb$_5$ (RbV$_3$Sb$_5$) are consistent with first-principles calculations that examine the evolution of their phonon spectra with pressure \cite{du2022superconductivity}. At zero pressure, the phonon spectra exhibit imaginary phonon modes at the $M$ and $L$ points, with the leading instability at the $L$ point. These phonon instabilities are $B_{1u}$ modes, corresponding to the CDW ground state at ambient pressure. When the pressure increases to 5 GPa (10 GPa), the imaginary modes in the phonon spectra disappear, indicating the suppression of CDW under pressure, and the hexagonal structure in Fig. \ref*{fig:fig1}(b) becomes the ground state. At higher pressures of 10 GPa (20 GPa), imaginary $B_{3g}$ phonon modes reappear around the $M$ point, suggesting a pressure-induced structural phase transition, consistent with the $\alpha$-$\beta$ transition in XRD measurements \cite{du2022superconductivity}.

The combination of electrical transport and XRD measurements in KV$_3$Sb$_5$ and RbV$_3$Sb$_5$ clearly reveals a significant modulation of superconductivity by pressure-induced structural phase transitions. The pressure that induces the first structural transition coincides with the boundary between the first and second superconducting domes, indicating a connection between the emergence of the second superconducting dome and the $\alpha$-$\beta$ structural transition. In KV$_3$Sb$_5$, the $\beta$-$\gamma$ structural transition occurs near the $T_{\rm c}$ maximum of the second superconducting dome, indicating that appearance of the $\gamma$ phase leads to the rapid suppression of SC above 20 GPa.

The pressure-tuning of SC and crystal structure beyond the full suppression of CDW have also been systematically studied for CsV$_3$Sb$_5$ \cite{zhao2021nodal,zhang2021pressure,chen2021highly,zhu2022double,yu2022pressure,tsirlin2022role}. Similar to KV$_3$Sb$_5$ and RbV$_3$Sb$_5$, CsV$_3$Sb$_5$ also exhibits a superconducting dome under high pressures (Fig.~\ref{highpPD}(c)), although XRD measurements indicate that its hexagonal structure persists up to at least 80 GPa, without a pressure-induced structural transition (Figs.\ref{XRD}(c) and (f)). Instead, the boundary separating the superconducting domes in CsV$_3$Sb$_5$ ($\sim15$~GPa) is associated with a crossover of the crystal structure from two-dimensional to three-dimensional, resulting from the formation of interlayer Sb-Sb bonds \cite{yu2022pressure,tsirlin2022role}. Such a change in structure is also evidenced in Raman scattering measurements, with significant changes in the intensities of $E_{2g}$ and $A_{1g}$ phonon modes above $\sim15$~GPa \cite{chen2021highly}. It is interesting to note that  non-hydrostatic pressure causes a hexagonal to monoclinic transition in CsV$_3$Sb$_5$, although the primary factor determining the evolution of SC with pressure remains changes in the electronic structure that result from Sb-Sb bonding \cite{tsirlin2023effect}. 

These high-pressure studies of $A$V$_3$Sb$_5$ indicate that the evolution of SC beyond full suppression of CDW is dominated by structural modulations, which manifests as structural phase transitions in KV$_3$Sb$_5$ and RbV$_3$Sb$_5$, and as a dimensionality crossover via interlayer Sb-Sb bonding in CsV$_3$Sb$_5$. Although the full crystal structures of the $\beta$  and $\gamma$ phases in KV$_3$Sb$_5$ and RbV$_3$Sb$_5$ have not been determined, they are likely three-dimensional structures, based on the evolution of other highly two-dimensional  systems under pressure \cite{du2022superconductivity}. Therefore, pressure-driven changes in the crystal structure, in particular the change from two-dimensional to three-dimensional structures, plays a dominant role in the modulations of SC in $A$V$_3$Sb$_5$ without CDW.  

\section{Tuning of CDW and SC in $\mathrm{CsV_3Sb_5}$ thin flakes}

The structure of $A$V$_3$Sb$_5$ is highly 2D, with V-Sb sheets separated by layers of alkali metal ions, which makes single crystals of $A$V$_3$Sb$_5$  mechanically exfoliable into thin flakes. Such a dimensional reduction is expected to weaken interlayer couplings and enhance quantum fluctuations and correlations, providing a route to tune the interplay between SC and CDW.

\begin{figure*}[htbp]
  \centering
  \includegraphics[width=2\columnwidth]{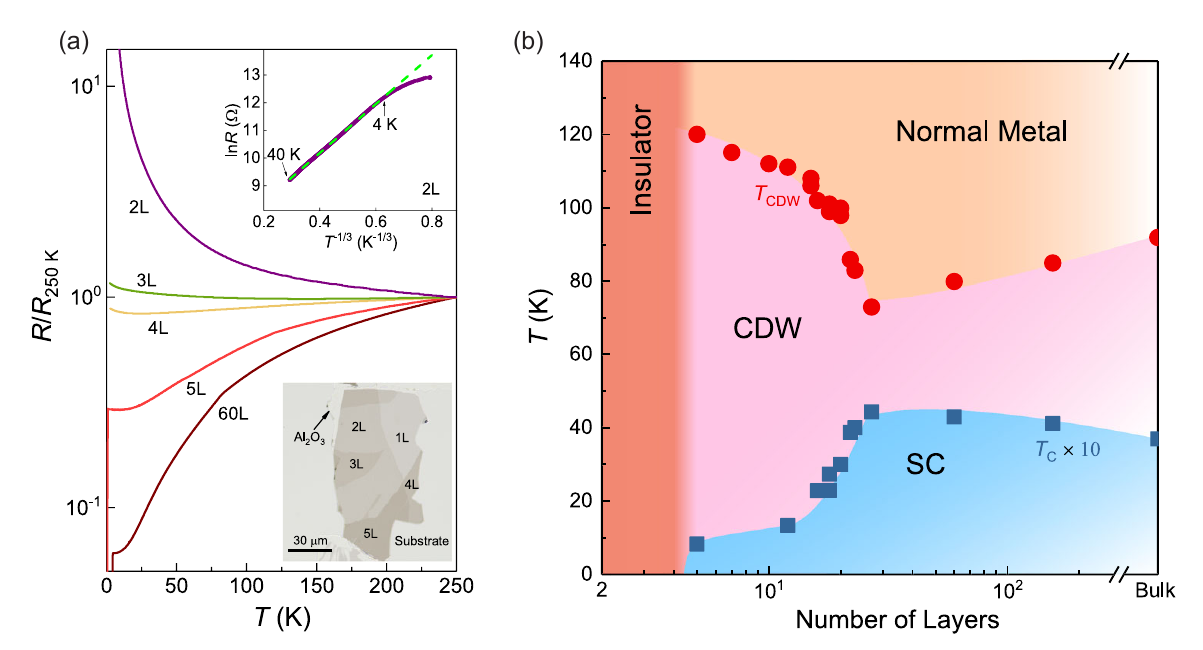}
  \caption{
    \label{Thickness}
     (a) Resistance of $\mathrm{CsV_3Sb_5}$ thin flakes as a function of temperature, for flakes ranging from 2~L to 60~L, normalized by the resistance at 250 K. The upper inset depicts the two-dimensional variable-range-hopping behavior in the 2~L sample. The lower inset presents an image of 1~L to 5~L flakes exfoliated onto an $\mathrm{Al_2O_3}$ substrate.
     (b) Thickness-temperature phase diagram of $\mathrm{CsV_3Sb_5}$. This figure is adapted from References \cite{song2023anomalous}. 
    }
\end{figure*}

Systematic experimental measurements on CsV$_3$Sb$_5$ thin flakes showed that the CDW order has an unusual non-monotonic thickness dependence \cite{song2023anomalous}. When the thickness of CsV$_3$Sb$_5$ is above $\sim$25 V-Sb layers (25~L), $T_{\rm CDW}$ decreases with decreasing thickness, consistent with reduced interlayer couplings weakening the CDW. Below 25~L, $T_{\rm CDW}$ increases dramatically with decreasing thickness and reaches 120~K, surpassing its bulk value of 94 K (Fig. \ref{Thickness} (b) ). In contrast, upon decreasing the flake thickness, $T_{\rm c}$  first increases slightly then decreases (Fig. \ref{Thickness} (b)), consistent with the notion that SC competes with CDW ($T_{\rm CDW}$ and $T_{\rm c}$ are anti-correlated, as shown in Fig. \ref{Thickness} (b)). The anomalous enhancement of $T_{\rm CDW}$ suggests a crossover of the CDW from being phonon-driven to being dominantly electron-driven, as the flake's thickness is reduced below 25~L \cite{song2023anomalous}. When the thickness is further reduced, a prominent metal-insulator transition is observed between 2~L and 5~L, suggesting that dimensional reduction leads to considerable electronic correlations in CsV$_3$Sb$_5$ thin flakes (Fig. \ref{Thickness} (a)). A non-saturating linear magnetoresistance has also been reported for CsV$_3$Sb$_5$ flakes below 20~nm ($\sim21$~L), which were suggested to result from 2D CDW fluctuations \cite{wei2022linear}.

\begin{figure}[htbp]
  \centering
  \includegraphics[width=\columnwidth]{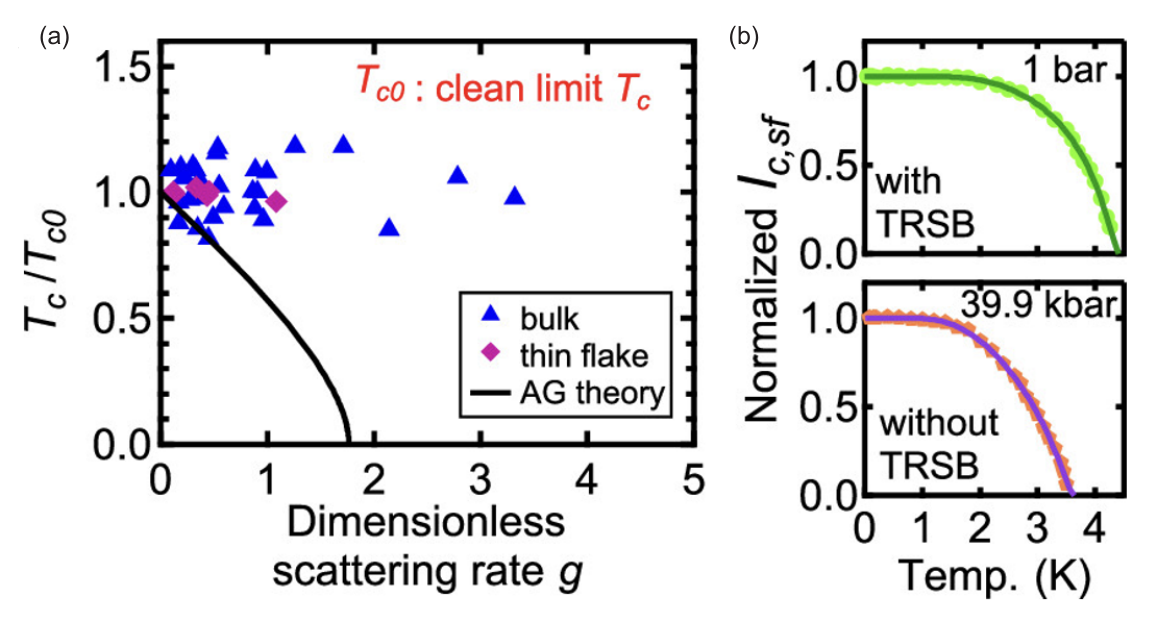}
  \caption{
    \label{pairing}
    (a) $T_{\rm c}$ of $\mathrm{CsV_3Sb_5}$ bulk and thin flakes, as a function of impurity scattering rate. The solid line denotes the expected $T_{\rm c}$ suppression according to Abrikosov-Gor'kov (AG) theory for sign-changing SC. (b) The temperature dependence of the normalized self-field critical current ($I_{c,sf}(T)/I_{c,sf}(0)$) at ambient pressure (1 bar, upper panel) and at 39.9 kbar (lower panel). The solid curves represent fits to the $s$-wave gap model \cite{zhang2023nodeless}.
  }
\end{figure}

Recently, it was proposed that the superfluid density can be probed via the self-field critical current ($I_{c,sf}$) in thin flake samples  \cite{talantsev2015universal}. Using this method and in combination with hydrostatic pressure, the $I_{c,sf}$ of CsV$_3$Sb$_5$ thin flakes ($\sim180$~L) were studied \cite{seo2015controlling,zhang2023nodeless}, which allows the pairing symmetry of CsV$_3$Sb$_5$ to be inferred. As shown in Fig. \ref{pairing} (b), the temperature dependence of normalized $I_{c,sf}$ at ambient pressure and at 39.9~kbar are both well described by an $s$-wave gap model. This shows that similar to bulk samples, the SC in $\mathrm{CsV_3Sb_5}$ thin flakes is nodeless and sign-preserving, regardless of whether CDW is present (ambient pressure) or not (39.9~kbar). This conclusion is further corroborated by the evolution of $T_{\rm c}$ in $\mathrm{CsV_3Sb_5}$ thin flakes with different impurity scattering rates (inferred from the residual resistivity), as shown in Fig. \ref{pairing} (a)). For a superconductor that has a sign-changing order parameter, Abrikosov-Gor'kov theory predicts that $T_{\rm c}$ is quickly suppressed by nonmagnetic impurities (solid line in Fig.~\ref{pairing}(a)). The observation of robust SC in both $\mathrm{CsV_3Sb_5}$ bulk and thin flakes evidence sign-preserving SC without symmetry enforced nodes. Further studies that examine the superconducting pairing symmetry in thinner $\mathrm{CsV_3Sb_5}$ thin flakes, particularly those below 25~L, will be instructive for understanding the thickness tuning of SC in $\mathrm{CsV_3Sb_5}$.     

\begin{figure}[htbp]
  \centering
  \includegraphics[width=\columnwidth]{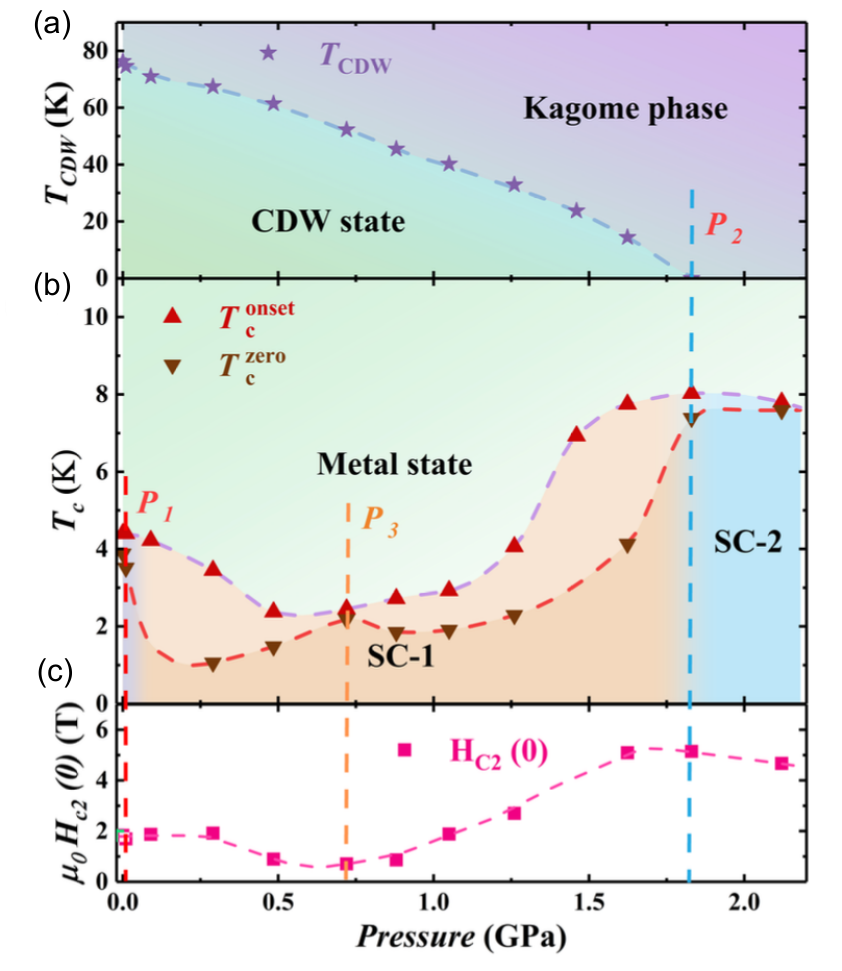}
  \caption{
    \label{pressure_thinflakes}
    The pressure-temperature phase diagram of 64~L $\mathrm{CsV_3Sb_5}$ thin flakes \cite{ye2024distinct}. (a) Pressure dependence of $T_{\rm CDW}$. The CDW is monotonically suppressed with increasing pressure and vanishes at $P_2$. (b) Pressure dependence of $T^{\rm zero}_{\rm c}$ and $T^{\rm onset}_{\rm c}$. (c) The upper critical field $\mu_0H_{c2}(0)$, as a function of pressure.
  }
\end{figure}

To shed light on the anomalous thickness dependence of  $T_{\rm CDW}$ and  $T_{\rm c}$ in CsV$_3$Sb$_5$ flakes (Fig. \ref{Thickness} (b)), electrical transport measurements under hydrostatic pressure up to 2.12 GPa were performed on $\approx$ 64 L CsV$_3$Sb$_5$ flakes\cite{ye2024distinct}. As in bulk CsV$_3$Sb$_5$ (Fig. \ref{lowpPD} (d)), the $P$-$T$ phase diagram can be obtained for the thin flake, with the evolution of CDW and SC delineated in Fig. \ref{pressure_thinflakes}. In bulk CsV$_3$Sb$_5$, $T_{\rm CDW}$ continuously decreases with increasing pressure until becoming fully suppressed around 2.0~GPa, and $T_{\rm c}$ exhibits two maximum at $P_1\sim$0.7 GPa and $P_2\sim$2.0 GPa \cite{chen2021double,yu2021unusual}. In the CsV$_3$Sb$_5$ thin flake, $T_{\rm CDW}$ also decreases monotonically with pressure and disappears at $P_2$ $\sim$ 1.8 GPa, similar to $P_2$ $\sim$ 2.0 GPa in the bulk. The pressure dependence of $T_{\rm c}$ ( $T_c^{\mathrm{onset}}$ and $T_c^{\mathrm{zero}}$) in the 64~L CsV$_3$Sb$_5$ thin flakes exhibits two SC domes, similar to bulk CsV$_3$Sb$_5$. However, whereas $P_2$ is similar in the thin flake and in the bulk, the first  $T_{\rm c}$ maximum is around ambient pressure ( $P_1<0.1$~GPa) for the thin flake, in contrast to $P_1\sim$0.7 GPa for the bulk. $T_{\rm c}^{\mathrm{onset}}$ and $T_{\rm c}^{\mathrm{zero}}$ converge at $P_1$ and $P_2$, consistent with results for the bulk \cite{chen2021double,yu2021unusual}. Between $P_1$ and $P_2$, it is likely that CDW domains induce a spatially inhomogeneous SC, leading to broad transitions similar to that in 1$T$-TaSe$_2$ \cite{joe2022emergence}. It is interesting to note at $P_3\sim0.7$~GPa,  $T_{\rm c}^{\mathrm{onset}}$ and $T_{\rm c}^{\mathrm{zero}}$ are also similar in the thin flake, whereas a similar behavior has not been reported in the bulk. The two dome SC in CsV$_3$Sb$_5$ thin flakes is corroborated by measurements of the upper critical field $\mu_0H_{c2}(0)$ as a function of pressure (Fig. \ref{pressure_thinflakes} (c)).    

These observations provide a perspective for understanding the unusual thickness dependence of  $T_{\rm c}$ in Fig. \ref{Thickness} (d):  reducing the number of layers moves $P_1$ from 0.7~GPa in the bulk to lower pressures, leading to an enhancement in $T_{\rm c}$. Below 25~L, $P_1$ becomes negative and a further reduction in the number of layers results in a decrease in $T_{\rm c}$, as $P_1$ is pushed further away from ambient pressure towards more negative pressures.

Collectively, the studies on $\mathrm{CsV_3Sb_5}$ thin flakes described above demonstrate thickness to be a powerful tuning parameter for unraveling the physics in kagome metals. Additional studies that tune the electron filling  in monolayer \cite{kim2023monolayer} or multilayer $A$V$_3$Sb$_5$, possibly in combination with hydrostatic pressure, will further shed light on the interplay between CDW and SC in $A$V$_3$Sb$_5$.

\section{Summary and outlook}

In this article we reviewed recent studies on the superconducting pairing and pressure/thickness tuning of $A$V$_3$Sb$_5$ systems. Although the research on the $A$V$_3$Sb$_5$ systems is still at a young age, tremendous progress have been made in understanding the unusual physical properties of these kagome metals. Here, we highlight several unsettled issues that are integral in understanding the $A$V$_3$Sb$_5$ compounds. 

While different experimental techniques consistently find multigap nodeless SC in CsV$_3$Sb$_5$, which remains robust upon tuning by pressure and doping (Fig.~\ref{PCS_pressure} and Fig.~\ref{fig:fig_doping2}),  the superconducting coupling strength is sensitive to pressure or strain, and varies significantly in studies using different techniques (Table \ref{Table_gap}). Furthermore, the pairing symmetries in KV$_3$Sb$_5$ and RbV$_3$Sb$_5$ are reported to be nodal at ambient pressure (Fig.~\ref{fig5_R_K135}) \cite{guguchia2023tunable}, different from CsV$_3$Sb$_5$. The origin of varying superconducting coupling strengths reported in CsV$_3$Sb$_5$, and how to consistently understand the pairing symmetries in different members compounds of $A$V$_3$Sb$_5$, are outstanding problems that need to be addressed. In addition, the SC in $A$V$_3$Sb$_5$ displays unique properties including (1) a low superfluid density resembling unconventional superconductors\cite{mielke2022time}, (2) a spatially modulated pair-density wave state \cite{chen2021roton}, and (3) Majorana zero modes in magnetic vortices\cite{liang2021three}. These features should be accounted for in models of SC in the $A$V$_3$Sb$_5$ compounds.

Upon applying pressure,  the three $A$V$_3$Sb$_5$ systems exhibit similarities that include (1) a competition between SC and CDW,  (2) an absence of CDW quantum critical point, and (3) a nonmonotonic evolution of $T_{\rm c}$ due to structural modulations (likely 2D to 3D) well beyond the suppression of CDW. Despite these commonalities in pressurized $A$V$_3$Sb$_5$, the evolution of $T_{\rm CDW}$ and $T_{\rm c}$ among the three member compounds cannot be explained by chemical pressure. As the atomic radius increases from K to Rb and Cs, $T_{\text{CDW}}$ does not vary monotonically, but peaks in RbV$_3$Sb$_5$, whereas $T_{\rm c}$ in CsV$_3$Sb$_5$ is significantly higher compared to KV$_3$Sb$_5$ and RbV$_3$Sb$_5$. A prominent two-dome SC and the emergence of a distinct CDW in CsV$_3$Sb$_5$ at low pressures (Fig.~\ref{lowpPD}(d)) \cite{zheng2022emergent}, also differentiates CsV$_3$Sb$_5$ from its sister compounds. Optical conductivity measurements indicate stronger electronic correlations in CsV$_3$Sb$_5$ compared to KV$_3$Sb$_5$ and RbV$_3$Sb$_5$ \cite{zhou2023electronic}, which likely plays a key role in understanding its unique properties. Disentangling the role of electron correlations and lattice instabilities, and identifying the primary driving mechanism for various orders, are key issues that need to be definitively resolved in $A$V$_3$Sb$_5$. On these fronts, kagome compounds that are structurally similar to $A$V$_3$Sb$_5$ but without intertwined orders may provide insights. The recently discovered $A$Ti$_3$Bi$_5$ ($A$=Cs,Rb) compounds is one such example, these compounds are kagome metals without CDWs, and SC appears upon applying pressure or chemical doping\cite{yi2023superconducting}. Spectroscopic signatures of unconventional electronic states, such as nematicity, have been observed in $A$Ti$_3$Bi$_5$\cite{li2023electronic}, motivating further detailed.

The suppression of $T_{\rm CDW}$ under pressure can be understood in terms of the electronic structure: DFT calculations show that pressure causes the vHs, which play an essential role in driving the CDW, to move away from the Fermi level,  and thus suppressing the CDW \cite{labollita2021tuning}. Applying a similar reasoning to hole-doped $A$V$_3$Sb$_5$ suggest that since the vHs move closer to the Fermi level, $T_{\rm CDW}$ should be enhanced, which is the opposite to what is seen experimentally. This discrepancy suggest that in the doped $A$V$_3$Sb$_5$ systems, disorder effects play an important role in determining properties of the CDW and SC, in addition to shifts of the Fermi level due to doped carriers.

Thickness is demonstrated to be a powerful tuning parameter for studying the interplay between SC and CDW in CsV$_3$Sb$_5$ thin flakes, revealing an unusual enhancement of $T_{\rm CDW}$ when the number of V-Sb layers is reduced below a threshold value of 25~L. Such an anomalous enhancement of $T_{\rm CDW}$ is accompanied by a weakening of SC, and is linked to the first SC dome in the $P$-$T$ phase diagram moving from $P_1\sim0.7$~GPa in the bulk towards lower and negative pressures in thin flakes. The intersection of dimensional reduction, vHs, and topological flat bands in exfoliable kagome metals, as exemplified by $A$V$_3$Sb$_5$, presents a rich frontier with the potential to uncover nontrivial physical states and properties. 

The vanadium-based kagome $A$V$_3$Sb$_5$ compounds offer a rich platform for studying CDW, SC, topological electronic states, electronic correlations, and their interplay. After a few years of research, there is now a clear understanding of the superconducting pairing symmetry in CsV$_3$Sb$_5$, and the interplay between CDW and SC upon pressure tuning in all three member compounds. There are also many unanswered questions in these fascinating materials, and through the process of addressing these issues, our understanding of quantum materials more broadly will also move forward.

\section*{Acknowledgements}
This work was supported by the National Key R$\&$D Program of China (Grant No.~2022YFA1402200), the Key R$\&$D Program of Zhejiang Province, China (Grant No.~2021C01002), the National Natural Science Foundation of China (Grants No.~11974306, No.~12034017), and the Zhejiang Provincial Natural Science Foundation of China (Grant No.~LR22A040002).

\section*{References}

\bibliographystyle{unsrt}
\bibliography{refs}

\begin{thebibliography}{100}

\bibitem{norman2016colloquium}
MR~Norman.
\newblock Colloquium: Herbertsmithite and the search for the quantum spin liquid.
\newblock {\em Reviews of Modern Physics}, 88(4):041002, 2016.

\bibitem{zhou2017quantum}
Yi~Zhou, Kazushi Kanoda, and Tai-Kai Ng.
\newblock Quantum spin liquid states.
\newblock {\em Reviews of Modern Physics}, 89(2):025003, 2017.

\bibitem{balents2010spin}
Leon Balents.
\newblock Spin liquids in frustrated magnets.
\newblock {\em Nature}, 464(7286):199--208, 2010.

\bibitem{yin2022topological}
Jia-Xin Yin, Biao Lian, and M~Zahid Hasan.
\newblock Topological kagome magnets and superconductors.
\newblock {\em Nature}, 612(7941):647--657, 2022.

\bibitem{ortiz2019new}
Brenden~R Ortiz, L{\'\i}dia~C Gomes, Jennifer~R Morey, Michal Winiarski, Mitchell Bordelon, John~S Mangum, Iain~WH Oswald, Jose~A Rodriguez-Rivera, James~R Neilson, Stephen~D Wilson, et~al.
\newblock New kagome prototype materials: discovery of {K}{V}$_3${Sb}$_5$, {Rb}{V}$_3${Sb}$_5$, and {Cs}{V}$_3${Sb}$_5$.
\newblock {\em Physical Review Materials}, 3(9):094407, 2019.

\bibitem{ortiz2020cs}
Brenden~R Ortiz, Samuel~ML Teicher, Yong Hu, Julia~L Zuo, Paul~M Sarte, Emily~C Schueller, AM~Milinda Abeykoon, Matthew~J Krogstad, Stephan Rosenkranz, Raymond Osborn, et~al.
\newblock {Cs}{V}$_3${Sb}$_5$: A {Z}$_2$ topological kagome metal with a superconducting ground state.
\newblock {\em Physical Review Letters}, 125(24):247002, 2020.

\bibitem{ortiz2021superconductivity}
Brenden~R Ortiz, Paul~M Sarte, Eric~M Kenney, Michael~J Graf, Samuel~ML Teicher, Ram Seshadri, and Stephen~D Wilson.
\newblock Superconductivity in the {Z}$_2$ kagome metal {K}{V}$_3${Sb}$_5$.
\newblock {\em Physical Review Materials}, 5(3):034801, 2021.

\bibitem{yin2021superconductivity}
Qiangwei Yin, Zhijun Tu, Chunsheng Gong, Yang Fu, Shaohua Yan, and Hechang Lei.
\newblock Superconductivity and normal-state properties of kagome metal {Rb}{V}$_3${Sb}$_5$ single crystals.
\newblock {\em Chinese Physics Letters}, 38(3):037403, 2021.

\bibitem{ortiz2021fermi}
Brenden~R Ortiz, Samuel~ML Teicher, Linus Kautzsch, Paul~M Sarte, Noah Ratcliff, John Harter, Jacob~PC Ruff, Ram Seshadri, and Stephen~D Wilson.
\newblock Fermi surface mapping and the nature of charge-density-wave order in the kagome superconductor {Cs}{V}$_3${Sb}$_5$.
\newblock {\em Physical Review X}, 11(4):041030, 2021.

\bibitem{lou2022charge}
Rui Lou, Alexander Fedorov, Qiangwei Yin, Andrii Kuibarov, Zhijun Tu, Chunsheng Gong, Eike~F Schwier, Bernd B{\"u}chner, Hechang Lei, and Sergey Borisenko.
\newblock Charge-density-wave-induced peak-dip-hump structure and the multiband superconductivity in a kagome superconductor {Cs}{V}$_3${Sb}$_5$.
\newblock {\em Physical Review Letters}, 128(3):036402, 2022.

\bibitem{cho2021emergence}
Soohyun Cho, Haiyang Ma, Wei Xia, Yichen Yang, Zhengtai Liu, Zhe Huang, Zhicheng Jiang, Xiangle Lu, Jishan Liu, Zhonghao Liu, et~al.
\newblock Emergence of new van hove singularities in the charge density wave state of a topological kagome metal {Rb}{V}$_3${Sb}$_5$.
\newblock {\em Physical Review Letters}, 127(23):236401, 2021.

\bibitem{nakayama2021multiple}
Kosuke Nakayama, Yongkai Li, Takemi Kato, Min Liu, Zhiwei Wang, Takashi Takahashi, Yugui Yao, and Takafumi Sato.
\newblock Multiple energy scales and anisotropic energy gap in the charge-density-wave phase of the kagome superconductor {Cs}{V}$_3${Sb}$_5$.
\newblock {\em Physical Review B}, 104(16):L161112, 2021.

\bibitem{liu2021charge}
Zhonghao Liu, Ningning Zhao, Qiangwei Yin, Chunsheng Gong, Zhijun Tu, Man Li, Wenhua Song, Zhengtai Liu, Dawei Shen, Yaobo Huang, et~al.
\newblock Charge-density-wave-induced bands renormalization and energy gaps in a kagome superconductor {Rb}{V}$_3${Sb}$_5$.
\newblock {\em Physical Review X}, 11(4):041010, 2021.

\bibitem{luo2022electronic}
Hailan Luo, Qiang Gao, Hongxiong Liu, Yuhao Gu, Dingsong Wu, Changjiang Yi, Junjie Jia, Shilong Wu, Xiangyu Luo, Yu~Xu, et~al.
\newblock Electronic nature of charge density wave and electron-phonon coupling in kagome superconductor {K}{V}$_3${Sb}$_5$.
\newblock {\em Nature communications}, 13(1):273, 2022.

\bibitem{kang2022twofold}
Mingu Kang, Shiang Fang, Jeong-Kyu Kim, Brenden~R Ortiz, Sae~Hee Ryu, Jimin Kim, Jonggyu Yoo, Giorgio Sangiovanni, Domenico Di~Sante, Byeong-Gyu Park, et~al.
\newblock Twofold van hove singularity and origin of charge order in topological kagome superconductor {Cs}{V}$_3${Sb}$_5$.
\newblock {\em Nature Physics}, 18(3):301--308, 2022.

\bibitem{hu2022topological}
Yong Hu, Samuel~ML Teicher, Brenden~R Ortiz, Yang Luo, Shuting Peng, Linwei Huai, Junzhang Ma, Nicholas~C Plumb, Stephen~D Wilson, Junfeng He, et~al.
\newblock Topological surface states and flat bands in the kagome superconductor {Cs}{V}$_3${Sb}$_5$.
\newblock {\em Science Bulletin}, 67(5):495--500, 2022.

\bibitem{hu2022coexistence}
Yong Hu, Xianxin Wu, Brenden~R Ortiz, Xinloong Han, Nicholas~C Plumb, Stephen~D Wilson, Andreas~P Schnyder, and Ming Shi.
\newblock Coexistence of trihexagonal and star-of-{D}avid pattern in the charge density wave of the kagome superconductor ${A}${V}$_3${Sb}$_5$.
\newblock {\em Physical Review B}, 106(24):L241106, 2022.

\bibitem{stahl2022temperature}
Q~Stahl, D~Chen, T~Ritschel, C~Shekhar, E~Sadrollahi, MC~Rahn, O~Ivashko, M~v Zimmermann, C~Felser, and J~Geck.
\newblock Temperature-driven reorganization of electronic order in {CsV}$_3${Sb}$_5$.
\newblock {\em Physical Review B}, 105(19):195136, 2022.

\bibitem{xiao2023coexistence}
Qian Xiao, Yihao Lin, Qizhi Li, Xiquan Zheng, Sonia Francoual, Christian Plueckthun, Wei Xia, Qingzheng Qiu, Shilong Zhang, Yanfeng Guo, et~al.
\newblock Coexistence of multiple stacking charge density waves in kagome superconductor {Cs}{V}$_3${Sb}$_5$.
\newblock {\em Physical Review Research}, 5(1):L012032, 2023.

\bibitem{kautzsch2023structural}
Linus Kautzsch, Brenden~R Ortiz, Krishnanand Mallayya, Jayden Plumb, Ganesh Pokharel, Jacob~PC Ruff, Zahirul Islam, Eun-Ah Kim, Ram Seshadri, and Stephen~D Wilson.
\newblock Structural evolution of the kagome superconductors ${A}${V}$_3${Sb}$_5$ ({A} = {K}, {Rb}, {Cs}) through charge density wave order.
\newblock {\em Physical Review Materials}, 7(2):024806, 2023.

\bibitem{luo2022posiible}
J.~Luo, Z.~Zhao, Y.~Z. Zhou, J.~Yang, A.~F. Fang, H.~T. Yang, H.~J. Gao, R.~Zhou, and Guo-qing Zheng.
\newblock Possible star-of-{D}avid pattern charge density wave with additional modulation in the kagome superconductor {CsV}$_3${Sb}$_5$.
\newblock {\em npj Quantum Materials}, 7(1):30, 2022.

\bibitem{frassineti2023microscopic}
Jonathan Frassineti, Pietro Bonf{\`a}, Giuseppe Allodi, Erick Garcia, Rong Cong, Brenden~R Ortiz, Stephen~D Wilson, Roberto De~Renzi, Vesna~F Mitrovi{\'c}, and Samuele Sanna.
\newblock Microscopic nature of the charge-density wave in the kagome superconductor {Rb}{V}$_3${Sb}$_5$.
\newblock {\em Physical Review Research}, 5(1):L012017, 2023.

\bibitem{yu2021concurrence}
FH~Yu, T~Wu, ZY~Wang, B~Lei, WZ~Zhuo, JJ~Ying, and XH~Chen.
\newblock Concurrence of anomalous hall effect and charge density wave in a superconducting topological kagome metal.
\newblock {\em Physical Review B}, 104(4):L041103, 2021.

\bibitem{jiang2021unconventional}
Yu-Xiao Jiang, Jia-Xin Yin, M~Michael Denner, Nana Shumiya, Brenden~R Ortiz, Gang Xu, Zurab Guguchia, Junyi He, Md~Shafayat Hossain, Xiaoxiong Liu, et~al.
\newblock Unconventional chiral charge order in kagome superconductor {K}{V}$_3${Sb}$_5$.
\newblock {\em Nature materials}, 20(10):1353--1357, 2021.

\bibitem{feng2021chiral}
Xilin Feng, Kun Jiang, Ziqiang Wang, and Jiangping Hu.
\newblock Chiral flux phase in the kagome superconductor ${A}${V}$_3${Sb}$_5$.
\newblock {\em Science bulletin}, 66(14):1384--1388, 2021.

\bibitem{shumiya2021intrinsic}
Nana Shumiya, Md~Shafayat Hossain, Jia-Xin Yin, Yu-Xiao Jiang, Brenden~R Ortiz, Hongxiong Liu, Youguo Shi, Qiangwei Yin, Hechang Lei, Songtian~S Zhang, et~al.
\newblock Intrinsic nature of chiral charge order in the kagome superconductor {Rb}{V}$_3${Sb}$_5$.
\newblock {\em Physical Review B}, 104(3):035131, 2021.

\bibitem{wang2021electronic}
Zhiwei Wang, Yu-Xiao Jiang, Jia-Xin Yin, Yongkai Li, Guan-Yong Wang, Hai-Li Huang, Sen Shao, Jinjin Liu, Peng Zhu, Nana Shumiya, et~al.
\newblock Electronic nature of chiral charge order in the kagome superconductor {Cs}{V}$_3${Sb}$_5$.
\newblock {\em Physical Review B}, 104(7):075148, 2021.

\bibitem{mielke2022time}
C~Mielke~III, Debarchan Das, J-X Yin, H~Liu, R~Gupta, Y-X Jiang, Marisa Medarde, X~Wu, Hechang~C Lei, J~Chang, et~al.
\newblock Time-reversal symmetry-breaking charge order in a kagome superconductor.
\newblock {\em Nature}, 602(7896):245--250, 2022.

\bibitem{khasanov2022time}
Rustem Khasanov, Debarchan Das, Ritu Gupta, Charles Mielke~III, Matthias Elender, Qiangwei Yin, Zhijun Tu, Chunsheng Gong, Hechang Lei, Ethan~T Ritz, et~al.
\newblock Time-reversal symmetry broken by charge order in {Cs}{V}$_3${Sb}$_5$.
\newblock {\em Physical Review Research}, 4(2):023244, 2022.

\bibitem{yu2021evidence}
Li~Yu, Chennan Wang, Yuhang Zhang, Mathias Sander, Shunli Ni, Zouyouwei Lu, Sheng Ma, Zhengguo Wang, Zhen Zhao, Hui Chen, et~al.
\newblock Evidence of a hidden flux phase in the topological kagome metal {Cs}{V}$_3${Sb}$_5$.
\newblock {\em arXiv preprint arXiv:2107.10714}, 2021.

\bibitem{hu2022time}
Yajian Hu, Soichiro Yamane, Giordano Mattoni, Kanae Yada, Keito Obata, Yongkai Li, Yugui Yao, Zhiwei Wang, Jingyuan Wang, Camron Farhang, et~al.
\newblock Time-reversal symmetry breaking in charge density wave of {Cs}{V}$_3${Sb}$_5$ detected by polar kerr effect.
\newblock {\em arXiv preprint arXiv:2208.08036}, 2022.

\bibitem{yang2020giant}
Shuo-Ying Yang, Yaojia Wang, Brenden~R Ortiz, Defa Liu, Jacob Gayles, Elena Derunova, Rafael Gonzalez-Hernandez, Libor {\v{S}}mejkal, Yulin Chen, Stuart~SP Parkin, et~al.
\newblock Giant, unconventional anomalous hall effect in the metallic frustrated magnet candidate, {K}{V}$_3${Sb}$_5$.
\newblock {\em Science advances}, 6(31):eabb6003, 2020.

\bibitem{zhou2022anomalous}
Xuebo Zhou, Hongxiong Liu, Wei Wu, Kun Jiang, Youguo Shi, Zheng Li, Yu~Sui, Jiangping Hu, and Jianlin Luo.
\newblock Anomalous thermal hall effect and anomalous nernst effect of {Cs}{V}$_3${Sb}$_5$.
\newblock {\em Physical Review B}, 105(20):205104, 2022.

\bibitem{chen2022anomalous}
Dong Chen, Bin He, Mengyu Yao, Yu~Pan, Haicheng Lin, Walter Schnelle, Yan Sun, Johannes Gooth, Louis Taillefer, and Claudia Felser.
\newblock Anomalous thermoelectric effects and quantum oscillations in the kagome metal {Cs}{V}$_3${Sb}$_5$.
\newblock {\em Physical Review B}, 105(20):L201109, 2022.

\bibitem{denner2021analysis}
M~Michael Denner, Ronny Thomale, and Titus Neupert.
\newblock Analysis of charge order in the kagome metal ${A}${V}$_3${Sb}$_5$ (${A}$ = {K, Rb, Cs}).
\newblock {\em Physical Review Letters}, 127(21):217601, 2021.

\bibitem{feng2021low}
Xilin Feng, Yi~Zhang, Kun Jiang, and Jiangping Hu.
\newblock Low-energy effective theory and symmetry classification of flux phases on the kagome lattice.
\newblock {\em Physical Review B}, 104(16):165136, 2021.

\bibitem{wu2022simultaneous}
Qiong Wu, ZX~Wang, QM~Liu, RS~Li, SX~Xu, QW~Yin, CS~Gong, ZJ~Tu, HC~Lei, T~Dong, et~al.
\newblock Simultaneous formation of two-fold rotation symmetry with charge order in the kagome superconductor {Cs}{V}$_3${Sb}$_5$ by optical polarization rotation measurement.
\newblock {\em Physical Review B}, 106(20):205109, 2022.

\bibitem{xu2022three}
Yishuai Xu, Zhuoliang Ni, Yizhou Liu, Brenden~R Ortiz, Qinwen Deng, Stephen~D Wilson, Binghai Yan, Leon Balents, and Liang Wu.
\newblock Three-state nematicity and magneto-optical kerr effect in the charge density waves in kagome superconductors.
\newblock {\em Nature physics}, 18(12):1470--1475, 2022.

\bibitem{asaba2023evidence}
T~Asaba, A~Onishi, Y~Kageyama, T~Kiyosue, K~Ohtsuka, S~Suetsugu, Y~Kohsaka, T~Gaggl, Y~Kasahara, H~Murayama, et~al.
\newblock Evidence for an odd-parity nematic phase above the charge density wave transition in kagome metal {Cs}{V}$_3${Sb}$_5$.
\newblock {\em arXiv preprint arXiv:2309.16985}, 2023.

\bibitem{lin2021complex}
Yu-Ping Lin and Rahul~M Nandkishore.
\newblock Complex charge density waves at van hove singularity on hexagonal lattices: Haldane-model phase diagram and potential realization in the kagome metals ${A}${V}$_3${Sb}$_5$ (${A}$ = {K, Rb, Cs}).
\newblock {\em Physical Review B}, 104(4):045122, 2021.

\bibitem{nie2022charge}
Linpeng Nie, Kuanglv Sun, Wanru Ma, Dianwu Song, Lixuan Zheng, Zuowei Liang, Ping Wu, Fanghang Yu, Jian Li, Min Shan, et~al.
\newblock Charge-density-wave-driven electronic nematicity in a kagome superconductor.
\newblock {\em Nature}, 604(7904):59--64, 2022.

\bibitem{chen2021roton}
Hui Chen, Haitao Yang, Bin Hu, Zhen Zhao, Jie Yuan, Yuqing Xing, Guojian Qian, Zihao Huang, Geng Li, Yuhan Ye, et~al.
\newblock Roton pair density wave in a strong-coupling kagome superconductor.
\newblock {\em Nature}, 599(7884):222--228, 2021.

\bibitem{xiang2021twofold}
Ying Xiang, Qing Li, Yongkai Li, Wei Xie, Huan Yang, Zhiwei Wang, Yugui Yao, and Hai-Hu Wen.
\newblock Twofold symmetry of c-axis resistivity in topological kagome superconductor {Cs}{V}$_3${Sb}$_5$ with in-plane rotating magnetic field.
\newblock {\em Nature communications}, 12(1):6727, 2021.

\bibitem{wang2024two}
Shuo Wang, Jing-Zhi Fang, Ze-Nan Wu, Sirong Lu, Zhongming Wei, Zhiwei Wang, Wen Huang, Yugui Yao, Jia-Jie Yang, Tingyong Chen, et~al.
\newblock Two-fold symmetric superconductivity in the kagome superconductor {Rb}{V}$_3${Sb}$_5$.
\newblock {\em Communications Physics}, 7(1):15, 2024.

\bibitem{liang2021three}
Zuowei Liang, Xingyuan Hou, Fan Zhang, Wanru Ma, Ping Wu, Zongyuan Zhang, Fanghang Yu, J-J Ying, Kun Jiang, Lei Shan, et~al.
\newblock Three-dimensional charge density wave and surface-dependent vortex-core states in a kagome superconductor {Cs}{V}$_3${Sb}$_5$.
\newblock {\em Physical Review X}, 11(3):031026, 2021.

\bibitem{jiang2023kagome}
Kun Jiang, Tao Wu, Jia-Xin Yin, Zhenyu Wang, M~Zahid Hasan, Stephen~D Wilson, Xianhui Chen, and Jiangping Hu.
\newblock Kagome superconductors ${A}${V}$_3${Sb}$_5$ (${A}$ = {K, Rb, Cs}).
\newblock {\em National Science Review}, 10(2):nwac199, 2023.

\bibitem{chen2022superconductivity}
Hui Chen, Bin Hu, Yuhan Ye, Haitao Yang, and Hong-Jun Gao.
\newblock Superconductivity and unconventional density waves in vanadium-based kagome materials ${A}${V}$_3${Sb}$_5$.
\newblock {\em Chinese Physics B}, 31(9):097405, 2022.

\bibitem{neupert2022charge}
Titus Neupert, M~Michael Denner, Jia-Xin Yin, Ronny Thomale, and M~Zahid Hasan.
\newblock Charge order and superconductivity in kagome materials.
\newblock {\em Nature Physics}, 18(2):137--143, 2022.

\bibitem{mi2023electrical}
Xin-Run Mi, Kun-Ya Yang, Yu-Han Gan, Long Zhang, Ai-Feng Wang, Yi-Sheng Chai, Xiao-Yuan Zhou, and Ming-Quan He.
\newblock Electrical and thermal transport properties of kagome metals ${A}${V}$_3${Sb}$_5$ (${A}$ = {K, Rb, Cs}).
\newblock {\em Tungsten}, 5(3):300--316, 2023.

\bibitem{nguyen2022electronic}
Thanh Nguyen and Mingda Li.
\newblock Electronic properties of correlated kagom{\'e} metals ${A}${V}$_3${Sb}$_5$ (${A}$ = {K}, {Rb}, {Cs}): A perspective.
\newblock {\em Journal of Applied Physics}, 131(6), 2022.

\bibitem{wilson2024kagome}
Stephen~D Wilson and Brenden~R Ortiz.
\newblock ${A}${V}$_3${Sb}$_5$ kagome superconductors.
\newblock {\em Nature Reviews Materials}, 9:420--432, 2024.

\bibitem{zhang2023superconducting}
Dongting Zhang, Chufan Chen, Lichang Yin, Yan’En Huang, Fengrui Shi, Yi~Liu, Xiaofeng Xu, Huiqiu Yuan, and Xin Lu.
\newblock Superconducting gap evolution of kagome metal {Cs}{V}$_3${Sb}$_5$ under pressure.
\newblock {\em Science China Physics, Mechanics \& Astronomy}, 66(2):227411, 2023.

\bibitem{wu_2021}
Xianxin Wu, Tilman Schwemmer, Tobias M{\"u}ller, Armando Consiglio, Giorgio Sangiovanni, Domenico Di~Sante, Yasir Iqbal, Werner Hanke, Andreas~P Schnyder, M~Michael Denner, et~al.
\newblock Nature of unconventional pairing in the kagome superconductors ${A}${V}$_3${Sb}$_5$ (${A}$ = {K, Rb, Cs}).
\newblock {\em Physical review letters}, 127(17):177001, 2021.

\bibitem{kiesel_2012}
Maximilian~L Kiesel and Ronny Thomale.
\newblock Sublattice interference in the kagome hubbard model.
\newblock {\em Physical Review B}, 86(12):121105, 2012.

\bibitem{wang_2020}
Yaojia Wang, Shuoying Yang, Pranava~K Sivakumar, Brenden~R Ortiz, Samuel~ML Teicher, Heng Wu, Abhay~K Srivastava, Chirag Garg, Defa Liu, Stuart~SP Parkin, et~al.
\newblock Proximity-induced spin-triplet superconductivity and edge supercurrent in the topological kagome metal, {K}$_{1-x}${V}$_3${Sb}$_5$.
\newblock {\em arXiv preprint arXiv:2012.05898}, 2020.

\bibitem{zhao2021nodal}
CC~Zhao, LS~Wang, W~Xia, QW~Yin, JM~Ni, YY~Huang, CP~Tu, ZC~Tao, ZJ~Tu, CS~Gong, et~al.
\newblock Nodal superconductivity and superconducting domes in the topological kagome metal {Cs}{V}$_3${Sb}$_5$.
\newblock {\em arXiv preprint arXiv:2102.08356}, 2021.

\bibitem{duan2021nodeless}
Weiyin Duan, Zhiyong Nie, Shuaishuai Luo, Fanghang Yu, Brenden~R Ortiz, Lichang Yin, Hang Su, Feng Du, An~Wang, Ye~Chen, et~al.
\newblock Nodeless superconductivity in the kagome metal {Cs}{V}$_3${Sb}$_5$.
\newblock {\em Science China Physics, Mechanics \& Astronomy}, 64(10):107462, 2021.

\bibitem{roppongi2023bulk}
M~Roppongi, K~Ishihara, Y~Tanaka, K~Ogawa, K~Okada, S~Liu, K~Mukasa, Y~Mizukami, Y~Uwatoko, R~Grasset, et~al.
\newblock Bulk evidence of anisotropic s-wave pairing with no sign change in the kagome superconductor {Cs}{V}$_3${Sb}$_5$.
\newblock {\em Nature Communications}, 14(1):667, 2023.

\bibitem{shan2022muon}
Zhaoyang Shan, Pabitra~K Biswas, Sudeep~K Ghosh, T~Tula, Adrian~D Hillier, Devashibhai Adroja, Stephen Cottrell, Guang-Han Cao, Yi~Liu, Xiaofeng Xu, et~al.
\newblock Muon spin relaxation study of the layered kagome superconductor {Cs}{V}$_3${Sb}$_5$.
\newblock {\em Physical Review Research}, 4(3):033145, 2022.

\bibitem{gupta2022microscopic}
Ritu Gupta, Debarchan Das, Charles~Hillis Mielke~III, Zurab Guguchia, Toni Shiroka, Christopher Baines, Marek Bartkowiak, Hubertus Luetkens, Rustem Khasanov, Qiangwei Yin, et~al.
\newblock Microscopic evidence for anisotropic multigap superconductivity in the {Cs}{V}$_3${Sb}$_5$ kagome superconductor.
\newblock {\em npj Quantum Materials}, 7(1):49, 2022.

\bibitem{gupta2022two}
Ritu Gupta, Debarchan Das, Charles Mielke~III, Ethan Ritz, Fabian Hotz, Qiangwei Yin, Zhijun Tu, Chunsheng Gong, Hechang Lei, Turan Birol, et~al.
\newblock Two types of charge order in the superconducting kagome material {Cs}{V}$_3${Sb}$_5$.
\newblock {\em arXiv preprint arXiv:2203.05055}, 2022.

\bibitem{xu2021multiband}
Han-Shu Xu, Ya-Jun Yan, Ruotong Yin, Wei Xia, Shijie Fang, Ziyuan Chen, Yuanji Li, Wenqi Yang, Yanfeng Guo, and Dong-Lai Feng.
\newblock Multiband superconductivity with sign-preserving order parameter in kagome superconductor {Cs}{V}$_3${Sb}$_5$.
\newblock {\em Physical Review Letters}, 127(18):187004, 2021.

\bibitem{yin2021strain}
Lichang Yin, Dongting Zhang, Chufan Chen, Ge~Ye, Fanghang Yu, Brenden~R Ortiz, Shuaishuai Luo, Weiyin Duan, Hang Su, Jianjun Ying, et~al.
\newblock Strain-sensitive superconductivity in the kagome metals {K}{V}$_3${Sb}$_5$ and {Cs}{V}$_3${Sb}$_5$ probed by point-contact spectroscopy.
\newblock {\em Physical Review B}, 104(17):174507, 2021.

\bibitem{he2022strong}
Ming-chong He, Hai Zi, Hong-xing Zhan, Yu-qing Zhao, Cong Ren, Xing-yuan Hou, Lei Shan, Qiang-hua Wang, Qiangwei Yin, Zhijun Tu, et~al.
\newblock Strong-coupling superconductivity in the kagome metal {Cs}{V}$_3${Sb}$_5$ revealed by soft point-contact spectroscopy.
\newblock {\em Physical Review B}, 106(10):104510, 2022.

\bibitem{mu2021s}
Chao Mu, Qiangwei Yin, Zhijun Tu, Chunsheng Gong, Hechang Lei, Zheng Li, and Jianlin Luo.
\newblock S-wave superconductivity in kagome metal {Cs}{V}$_3${Sb}$_5$ revealed by $^{121/123}${Sb} {NQR} and $^{51}${V} {NMR} measurements.
\newblock {\em Chinese Physics Letters}, 38(7):077402, 2021.

\bibitem{zhong2023nodeless}
Yigui Zhong, Jinjin Liu, Xianxin Wu, Zurab Guguchia, J-X Yin, Akifumi Mine, Yongkai Li, Sahand Najafzadeh, Debarchan Das, Charles Mielke~III, et~al.
\newblock Nodeless electron pairing in {Cs}{V}$_3${Sb}$_5$-derived kagome superconductors.
\newblock {\em Nature}, 617(7961):488--492, 2023.

\bibitem{mine2024direct}
Akifumi Mine, Yigui Zhong, Jinjin Liu, Takeshi Suzuki, Sahand Najafzadeh, Takumi Uchiyama, Jia-Xin Yin, Xianxin Wu, Xun Shi, Zhiwei Wang, et~al.
\newblock Direct observation of anisotropic cooper pairing in kagome superconductor {Cs}{V}$_3${Sb}$_5$.
\newblock {\em arXiv preprint arXiv:2404.18472}, 2024.

\bibitem{van_1975}
Craig~T Van~Degrift.
\newblock Tunnel diode oscillator for 0.001 ppm measurements at low temperatures.
\newblock {\em Review of Scientific Instruments}, 46(5):599--607, 1975.

\bibitem{zhang2023nodeless}
Wei Zhang, Xinyou Liu, Lingfei Wang, Chun~Wai Tsang, Zheyu Wang, Siu~Tung Lam, Wenyan Wang, Jianyu Xie, Xuefeng Zhou, Yusheng Zhao, et~al.
\newblock Nodeless superconductivity in kagome metal {Cs}{V}$_3${Sb}$_5$ with and without time reversal symmetry breaking.
\newblock {\em Nano Letters}, 23(3):872--879, 2023.

\bibitem{balatsky_2006}
Alexander~V Balatsky, Ilya Vekhter, and Jian-Xin Zhu.
\newblock Impurity-induced states in conventional and unconventional superconductors.
\newblock {\em Reviews of Modern Physics}, 78(2):373, 2006.

\bibitem{anderson_1959}
Philip~W Anderson.
\newblock Theory of dirty superconductors.
\newblock {\em Journal of Physics and Chemistry of Solids}, 11(1-2):26--30, 1959.

\bibitem{prozorov_2006}
Ruslan Prozorov and Russell~W Giannetta.
\newblock Magnetic penetration depth in unconventional superconductors.
\newblock {\em Superconductor Science and Technology}, 19(8):R41, 2006.

\bibitem{prozorov2011london}
Ruslan Prozorov and Vladimir~G Kogan.
\newblock London penetration depth in iron-based superconductors.
\newblock {\em Reports on Progress in Physics}, 74(12):124505, 2011.

\bibitem{smidman2017superconductivity}
M~Smidman, MB~Salamon, HQ~Yuan, and DF~Agterberg.
\newblock Superconductivity and spin--orbit coupling in non-centrosymmetric materials: a review.
\newblock {\em Reports on Progress in Physics}, 80(3):036501, 2017.

\bibitem{le2022fermi}
Tian Le, LiQiang Che, Qi~Huang, Kevin Huang, ZhaoFeng Ding, Lei Shu, and Xin Lu.
\newblock Fermi surface gapping in the hidden order state of {Pr}{Fe}$_4${P}$_{12}$ by point-contact spectroscopy.
\newblock {\em Science China Physics, Mechanics \& Astronomy}, 65(3):237412, 2022.

\bibitem{chen2021double}
KY~Chen, NN~Wang, QW~Yin, YH~Gu, K~Jiang, ZJ~Tu, CS~Gong, Y~Uwatoko, JP~Sun, HC~Lei, et~al.
\newblock Double superconducting dome and triple enhancement of ${T}$$_c$ in the kagome superconductor {Cs}{V}$_3${Sb}$_5$ under high pressure.
\newblock {\em Physical Review Letters}, 126(24):247001, 2021.

\bibitem{yu2021unusual}
FH~Yu, DH~Ma, WZ~Zhuo, SQ~Liu, XK~Wen, Bin Lei, JJ~Ying, and XH~Chen.
\newblock Unusual competition of superconductivity and charge-density-wave state in a compressed topological kagome metal.
\newblock {\em Nature communications}, 12(1):3645, 2021.

\bibitem{qian2021revealing}
Tiema Qian, Morten~H Christensen, Chaowei Hu, Amartyajyoti Saha, Brian~M Andersen, Rafael~M Fernandes, Turan Birol, and Ni~Ni.
\newblock Revealing the competition between charge density wave and superconductivity in {Cs}{V}$_3${Sb}$_5$ through uniaxial strain.
\newblock {\em Physical Review B}, 104(14):144506, 2021.

\bibitem{yang2023plane}
Xiaoran Yang, Qi~Tang, Qiuyun Zhou, Huaiping Wang, Yi~Li, Xue Fu, Jiawen Zhang, Yu~Song, Huiqiu Yuan, Pengcheng Dai, et~al.
\newblock In-plane uniaxial-strain tuning of superconductivity and charge-density wave in {Cs}{V}$_3${Sb}$_5$.
\newblock {\em Chinese Physics B}, 32(12):127101, 2023.

\bibitem{guguchia2023tunable}
Zurab Guguchia, C~Mielke~III, Debarchan Das, Ritu Gupta, J-X Yin, Hongxiong Liu, Qiangwei Yin, Morten~Holm Christensen, Zhijun Tu, Chunsheng Gong, et~al.
\newblock Tunable unconventional kagome superconductivity in charge ordered {Rb}{V}$_3${Sb}$_5$ and {K}{V}$_3${Sb}$_5$.
\newblock {\em Nature communications}, 14(1):153, 2023.

\bibitem{hegger2000pressure}
H~Hegger, C~Petrovic, EG~Moshopoulou, MF~Hundley, JL~Sarrao, Z~Fisk, and JD~Thompson.
\newblock Pressure-induced superconductivity in quasi-2{D} {Ce}{Rh}{In}$_5$.
\newblock {\em Physical Review Letters}, 84(21):4986, 2000.

\bibitem{park2006hidden}
Tuson Park, F~Ronning, HQ~Yuan, MB~Salamon, R~Movshovich, JL~Sarrao, and JD~Thompson.
\newblock Hidden magnetism and quantum criticality in the heavy fermion superconductor {Ce}{Rh}{In}$_5$.
\newblock {\em Nature}, 440(7080):65--68, 2006.

\bibitem{yuan2003observation}
HQ~Yuan, FM~Grosche, M~Deppe, C~Geibel, G~Sparn, and F~Steglich.
\newblock Observation of two distinct superconducting phases in {Ce}{Cu}$_2${Si}$_2$.
\newblock {\em Science}, 302(5653):2104--2107, 2003.

\bibitem{weng2016multiple}
ZF~Weng, M~Smidman, L~Jiao, Xin Lu, and HQ~Yuan.
\newblock Multiple quantum phase transitions and superconductivity in {Ce}-based heavy fermions.
\newblock {\em Reports on Progress in Physics}, 79(9):094503, 2016.

\bibitem{shen2020strange}
Bin Shen, Yongjun Zhang, Yashar Komijani, Michael Nicklas, Robert Borth, An~Wang, Ye~Chen, Zhiyong Nie, Rui Li, Xin Lu, et~al.
\newblock Strange-metal behaviour in a pure ferromagnetic kondo lattice.
\newblock {\em Nature}, 579(7797):51--55, 2020.

\bibitem{gruner2017charge}
Thomas Gruner, Dongjin Jang, Zita Huesges, Raul Cardoso-Gil, Gerhard~H Fecher, Michael~M Koza, Oliver Stockert, Andrew~P Mackenzie, Manuel Brando, and Christoph Geibel.
\newblock Charge density wave quantum critical point with strong enhancement of superconductivity.
\newblock {\em Nature Physics}, 13(10):967--972, 2017.

\bibitem{goh2015ambient}
Swee~K Goh, DA~Tompsett, PJ~Saines, HC~Chang, T~Matsumoto, M~Imai, K~Yoshimura, and FM~Grosche.
\newblock Ambient pressure structural quantum critical point in the phase diagram of ({Ca}$_x${Sr}$_{1-x}$)$_3${Rh}$_4${Sn}$_{13}$.
\newblock {\em Physical review letters}, 114(9):097002, 2015.

\bibitem{gabovich2001charge}
AM~Gabovich, AI~Voitenko, JF~Annett, and M~Ausloos.
\newblock Charge-and spin-density-wave superconductors.
\newblock {\em Superconductor Science and Technology}, 14(4):R1, 2001.

\bibitem{du2020interplay}
F~Du, H~Su, SS~Luo, B~Shen, ZY~Nie, LC~Yin, Y~Chen, R~Li, M~Smidman, and HQ~Yuan.
\newblock Interplay between charge density wave order and superconductivity in {La}{Au}{Sb}$_2$ under pressure.
\newblock {\em Physical Review B}, 102(14):144510, 2020.

\bibitem{shen2020evolution}
B~Shen, F~Du, R~Li, A~Thamizhavel, M~Smidman, ZY~Nie, SS~Luo, T~Le, Z~Hossain, and HQ~Yuan.
\newblock Evolution of charge density wave order and superconductivity under pressure in {La}{Pt}$_2${Si}$_2$.
\newblock {\em Physical Review B}, 101(14):144501, 2020.

\bibitem{klintberg2012pressure}
Lina~E Klintberg, Swee~K Goh, Patricia~L Alireza, Paul~J Saines, David~A Tompsett, Peter~W Logg, Jinhu Yang, Bin Chen, Kazuyoshi Yoshimura, and F~Malte Grosche.
\newblock Pressure-induced and composition-induced structural quantum phase transition in the cubic superconductor ({Sr},{Ca})$_3${Ir}$_4${Sn}$_{13}$.
\newblock {\em arXiv preprint arXiv:1202.3282}, 2012.

\bibitem{mathur1998magnetically}
ND~Mathur, FM~Grosche, SR~Julian, IR~Walker, DM~Freye, RKW Haselwimmer, and GG~Lonzarich.
\newblock Magnetically mediated superconductivity in heavy fermion compounds.
\newblock {\em Nature}, 394(6688):39--43, 1998.

\bibitem{dai2009iron}
Jianhui Dai, Qimiao Si, Jian-Xin Zhu, and Elihu Abrahams.
\newblock Iron pnictides as a new setting for quantum criticality.
\newblock {\em Proceedings of the National Academy of Sciences}, 106(11):4118--4121, 2009.

\bibitem{shibauchi2014a}
T~Shibauchi, A~Carrington, and Y~Matsuda.
\newblock A quantum critical point lying beneath the superconducting dome in iron pnictides.
\newblock {\em Annu. Rev. Condens. Matter Phys.}, 5(1):113--135, 2014.

\bibitem{du2021pressure}
Feng Du, Shuaishuai Luo, Brenden~R Ortiz, Ye~Chen, Weiyin Duan, Dongting Zhang, Xin Lu, Stephen~D Wilson, Yu~Song, and Huiqiu Yuan.
\newblock Pressure-induced double superconducting domes and charge instability in the kagome metal {K}{V}$_3${Sb}$_5$.
\newblock {\em Physical Review B}, 103(22):L220504, 2021.

\bibitem{wang2021competition}
NN~Wang, KY~Chen, QW~Yin, YNN Ma, BY~Pan, X~Yang, XY~Ji, SL~Wu, PF~Shan, SX~Xu, et~al.
\newblock Competition between charge-density-wave and superconductivity in the kagome metal {Rb}{V}$_3${Sb}$_5$.
\newblock {\em Physical Review Research}, 3(4):043018, 2021.

\bibitem{du2022evolution}
Feng Du, Shuaishuai Luo, Rui Li, Brenden~R Ortiz, Ye~Chen, Stephen~D Wilson, Yu~Song, and Huiqiu Yuan.
\newblock Evolution of superconductivity and charge order in pressurized {Rb}{V}$_3${Sb}$_5$.
\newblock {\em Chinese Physics B}, 31(1):017404, 2022.

\bibitem{zhang2021first}
Jian-Feng Zhang, Kai Liu, and Zhong-Yi Lu.
\newblock First-principles study of the double-dome superconductivity in the kagome material {Cs}{V}$_3${Sb}$_5$ under pressure.
\newblock {\em Physical Review B}, 104(19):195130, 2021.

\bibitem{song2023anomalous}
Boqin Song, Tianping Ying, Xianxin Wu, Wei Xia, Qiangwei Yin, Qinghua Zhang, Yanpeng Song, Xiaofan Yang, Jiangang Guo, Lin Gu, et~al.
\newblock Anomalous enhancement of charge density wave in kagome superconductor {Cs}{V}$_3${Sb}$_5$ approaching the 2{D} limit.
\newblock {\em Nature Communications}, 14(1):2492, 2023.

\bibitem{hou2023effect}
J~Hou, KY~Chen, JP~Sun, Z~Zhao, YH~Zhang, PF~Shan, NN~Wang, H~Zhang, K~Zhu, Y~Uwatoko, et~al.
\newblock Effect of hydrostatic pressure on the unconventional charge density wave and superconducting properties in two distinct phases of doped kagome superconductors {Cs}{V}$_{3-x}${Ti}$_x${Sb}$_5$.
\newblock {\em Physical Review B}, 107(14):144502, 2023.

\bibitem{kang2023charge}
Mingu Kang, Shiang Fang, Jonggyu Yoo, Brenden~R Ortiz, Yuzki~M Oey, Jonghyeok Choi, Sae~Hee Ryu, Jimin Kim, Chris Jozwiak, Aaron Bostwick, et~al.
\newblock Charge order landscape and competition with superconductivity in kagome metals.
\newblock {\em Nature Materials}, 22(2):186--193, 2023.

\bibitem{consiglio2022van}
Armando Consiglio, Tilman Schwemmer, Xianxin Wu, Werner Hanke, Titus Neupert, Ronny Thomale, Giorgio Sangiovanni, and Domenico Di~Sante.
\newblock Van hove tuning of ${A}${V}$_3${Sb}$_5$ kagome metals under pressure and strain.
\newblock {\em Physical Review B}, 105(16):165146, 2022.

\bibitem{li2022discovery}
Haoxiang Li, Gilberto Fabbris, AH~Said, JP~Sun, Yu-Xiao Jiang, J-X Yin, Yun-Yi Pai, Sangmoon Yoon, Andrew~R Lupini, CS~Nelson, et~al.
\newblock Discovery of conjoined charge density waves in the kagome superconductor {Cs}{V}$_3${Sb}$_5$.
\newblock {\em Nature communications}, 13(1):6348, 2022.

\bibitem{oey2022tuning}
Yuzki~M Oey, Farnaz Kaboudvand, Brenden~R Ortiz, Ram Seshadri, and Stephen~D Wilson.
\newblock Tuning charge density wave order and superconductivity in the kagome metals {K}{V}$_3${Sb}$_{5-x}${Sn}$_x$ and {Rb}{V}$_3${Sb}$_{5-x}${Sn}$_x$.
\newblock {\em Physical Review Materials}, 6(7):074802, 2022.

\bibitem{song2021competition}
Yanpeng Song, Tianping Ying, Xu~Chen, Xu~Han, Xianxin Wu, Andreas~P Schnyder, Yuan Huang, Jian-gang Guo, and Xiaolong Chen.
\newblock Competition of superconductivity and charge density wave in selective oxidized {Cs}{V}$_3${Sb}$_5$ thin flakes.
\newblock {\em Physical review letters}, 127(23):237001, 2021.

\bibitem{yang2022titanium}
Haitao Yang, Zihao Huang, Yuhang Zhang, Zhen Zhao, Jinan Shi, Hailan Luo, Lin Zhao, Guojian Qian, Hengxin Tan, Bin Hu, et~al.
\newblock Titanium doped kagome superconductor {Cs}{V}$_{3-x}${Ti}$_x${Sb}$_5$ and two distinct phases.
\newblock {\em Science Bulletin}, 67(21):2176--2185, 2022.

\bibitem{lin2022multidome}
Yu-Ping Lin and Rahul~M Nandkishore.
\newblock Multidome superconductivity in charge density wave kagome metals.
\newblock {\em Physical Review B}, 106(6):L060507, 2022.

\bibitem{li2022tuning}
Yongkai Li, Qing Li, Xinwei Fan, Jinjin Liu, Qi~Feng, Min Liu, Chunlei Wang, Jia-Xin Yin, Junxi Duan, Xiang Li, et~al.
\newblock Tuning the competition between superconductivity and charge order in the kagome superconductor {Cs}({V}$_{1-x}${Nb}$_x$)$_3${Sb}$_5$.
\newblock {\em Physical Review B}, 105(18):L180507, 2022.

\bibitem{si2022charge}
Jian-Guo Si, Wen-Jian Lu, Yu-Ping Sun, Peng-Fei Liu, and Bao-Tian Wang.
\newblock Charge density wave and pressure-dependent superconductivity in the kagome metal {Cs}{V}$_3${Sb}$_5$: A first-principles study.
\newblock {\em Physical Review B}, 105(2):024517, 2022.

\bibitem{wang2022charge}
Chongze Wang, Shuyuan Liu, Hyunsoo Jeon, Yu~Jia, and Jun-Hyung Cho.
\newblock Charge density wave and superconductivity in the kagome metal {Cs}{V}$_3${Sb}$_5$ around a pressure-induced quantum critical point.
\newblock {\em Physical Review Materials}, 6(9):094801, 2022.

\bibitem{zheng2022emergent}
Lixuan Zheng, Zhimian Wu, Ye~Yang, Linpeng Nie, Min Shan, Kuanglv Sun, Dianwu Song, Fanghang Yu, Jian Li, Dan Zhao, et~al.
\newblock Emergent charge order in pressurized kagome superconductor {Cs}{V}$_3${Sb}$_5$.
\newblock {\em Nature}, 611(7937):682--687, 2022.

\bibitem{oey2022fermi}
Yuzki~M Oey, Brenden~R Ortiz, Farnaz Kaboudvand, Jonathan Frassineti, Erick Garcia, Rong Cong, Samuele Sanna, Vesna~F Mitrovi{\'c}, Ram Seshadri, and Stephen~D Wilson.
\newblock Fermi level tuning and double-dome superconductivity in the kagome metal {Cs}{V}$_3${Sb}$_{5-x}${Sn}$_x$.
\newblock {\em Physical Review Materials}, 6(4):L041801, 2022.

\bibitem{kautzsch2023incommensurate}
Linus Kautzsch, Yuzki~M Oey, Hong Li, Zheng Ren, Brenden~R Ortiz, Ganesh Pokharel, Ram Seshadri, Jacob Ruff, Terawit Kongruengkit, John~W Harter, et~al.
\newblock Incommensurate charge-stripe correlations in the kagome superconductor {Cs}{V}$_3${Sb}$_{5-x}${Sn}$_x$.
\newblock {\em npj Quantum Materials}, 8(1):37, 2023.

\bibitem{feng2023commensurate}
XY~Feng, Z~Zhao, J~Luo, J~Yang, AF~Fang, HT~Yang, HJ~Gao, R~Zhou, and Guo-qing Zheng.
\newblock Commensurate-to-incommensurate transition of charge-density-wave order and a possible quantum critical point in pressurized kagome metal {Cs}{V}$_3${Sb}$_5$.
\newblock {\em npj Quantum Materials}, 8(1):23, 2023.

\bibitem{stier2024pressure}
F~Stier, A-A Haghighirad, G~Garbarino, S~Mishra, N~Stilkerich, D~Chen, C~Shekhar, T~Lacmann, C~Felser, T~Ritschel, et~al.
\newblock Pressure-dependent electronic superlattice in the {K}agome-superconductor {Cs}{V}$_3${Sb}$_5$.
\newblock {\em arXiv preprint arXiv:2404.14790}, 2024.

\bibitem{sur2023optimized}
Yeahan Sur, Kwang-Tak Kim, Sukho Kim, and Kee~Hoon Kim.
\newblock Optimized superconductivity in the vicinity of a nematic quantum critical point in the kagome superconductor {Cs}({V}$_{1-x}${Ti}$_x$)$_3${Sb}$_5$.
\newblock {\em Nature Communications}, 14(1):3899, 2023.

\bibitem{liu2023doping}
Yixuan Liu, Yuan Wang, Yongqing Cai, Zhanyang Hao, Xiao-Ming Ma, Le~Wang, Cai Liu, Jian Chen, Liang Zhou, Jinhua Wang, et~al.
\newblock Doping evolution of superconductivity, charge order, and band topology in hole-doped topological kagome superconductors {Cs}({V}$_{1-x}${Ti}$_x$)$_3${Sb}$_5$.
\newblock {\em Physical Review Materials}, 7(6):064801, 2023.

\bibitem{liu2022evolution}
Mengqin Liu, Tao Han, Xinran Hu, Yubing Tu, Zongyuan Zhang, Mingsheng Long, Xingyuan Hou, Qingge Mu, and Lei Shan.
\newblock Evolution of superconductivity and charge density wave through {Ta} and {Mo} doping in $\mathrm{CsV}_{3}\mathrm{Sb}_{5}$.
\newblock {\em Physical Review B}, 106(14):L140501, 2022.

\bibitem{ding2022effect}
Gaofeng Ding, Hongliang Wo, Yiqing Gu, Yimeng Gu, and Jun Zhao.
\newblock Effect of chromium doping on superconductivity and charge density wave order in the kagome metal {Cs}({V}$_{1-x}${Cr}$_x$)$_3${Sb}$_5$.
\newblock {\em Physical Review B}, 106(23):235151, 2022.

\bibitem{wen2023emergent}
Xikai Wen, Fanghang Yu, Zhigang Gui, Yuqing Zhang, Xingyuan Hou, Lei Shan, Tao Wu, Ziji Xiang, Zhenyu Wang, Jianjun Ying, et~al.
\newblock Emergent superconducting fluctuations in compressed kagome superconductor {Cs}{V}$_3${Sb}$_5$.
\newblock {\em Science Bulletin}, 68(3):259--265, 2023.

\bibitem{ashcroft1968metallic}
Neil~W Ashcroft.
\newblock Metallic hydrogen: A high-temperature superconductor?
\newblock {\em Physical Review Letters}, 21(26):1748, 1968.

\bibitem{drozdov2019superconductivity}
AP~Drozdov, PP~Kong, VS~Minkov, SP~Besedin, MA~Kuzovnikov, S~Mozaffari, L~Balicas, FF~Balakirev, DE~Graf, VB~Prakapenka, et~al.
\newblock Superconductivity at 250{K} in lanthanum hydride under high pressures.
\newblock {\em Nature}, 569(7757):528--531, 2019.

\bibitem{somayazulu2019evidence}
Maddury Somayazulu, Muhtar Ahart, Ajay~K Mishra, Zachary~M Geballe, Maria Baldini, Yue Meng, Viktor~V Struzhkin, and Russell~J Hemley.
\newblock Evidence for superconductivity above 260{K} in lanthanum superhydride at megabar pressures.
\newblock {\em Physical review letters}, 122(2):027001, 2019.

\bibitem{yu2022pressure}
Fanghang Yu, Xudong Zhu, Xikai Wen, Zhigang Gui, Zeyu Li, Yulei Han, Tao Wu, Zhenyu Wang, Ziji Xiang, Zhenhua Qiao, et~al.
\newblock Pressure-induced dimensional crossover in a kagome superconductor.
\newblock {\em Physical Review Letters}, 128(7):077001, 2022.

\bibitem{du2022superconductivity}
Feng Du, Rui Li, Shuaishuai Luo, Yu~Gong, Yanchun Li, Sheng Jiang, Brenden~R Ortiz, Yi~Liu, Xiaofeng Xu, Stephen~D Wilson, et~al.
\newblock Superconductivity modulated by structural phase transitions in pressurized vanadium-based kagome metals.
\newblock {\em Physical Review B}, 106(2):024516, 2022.

\bibitem{zhang2021pressure}
Zhuyi Zhang, Zheng Chen, Ying Zhou, Yifang Yuan, Shuyang Wang, Jing Wang, Haiyang Yang, Chao An, Lili Zhang, Xiangde Zhu, et~al.
\newblock Pressure-induced reemergence of superconductivity in the topological kagome metal {Cs}{V}$_3${Sb}$_5$.
\newblock {\em Physical Review B}, 103(22):224513, 2021.

\bibitem{chen2021highly}
Xu~Chen, Xinhui Zhan, Xiaojun Wang, Jun Deng, Xiao-Bing Liu, Xin Chen, Jian-Gang Guo, and Xiaolong Chen.
\newblock Highly robust reentrant superconductivity in {Cs}{V}$_3${Sb}$_5$ under pressure.
\newblock {\em Chinese Physics Letters}, 38(5):057402, 2021.

\bibitem{zhu2022double}
CC~Zhu, XF~Yang, W~Xia, QW~Yin, LS~Wang, CC~Zhao, DZ~Dai, CP~Tu, BQ~Song, ZC~Tao, et~al.
\newblock Double-dome superconductivity under pressure in the {V}-based kagome metals ${A}${V}$_3${Sb}$_5$ (${A}$= {Rb} and {K}).
\newblock {\em Physical Review B}, 105(9):094507, 2022.

\bibitem{tsirlin2022role}
Alexander Tsirlin, Pierre Fertey, Brenden~R Ortiz, Berina Klis, Valentino Merkl, Martin Dressel, Stephen Wilson, and Ece Uykur.
\newblock Role of {Sb} in the superconducting kagome metal {Cs}{V}$_3${Sb}$_5$ revealed by its anisotropic compression.
\newblock {\em SciPost Physics}, 12(2):049, 2022.

\bibitem{tsirlin2023effect}
Alexander~A Tsirlin, Brenden~R Ortiz, Martin Dressel, Stephen~D Wilson, Stephan Winnerl, and Ece Uykur.
\newblock Effect of nonhydrostatic pressure on the superconducting kagome metal {Cs}{V}$_3${Sb}$_5$.
\newblock {\em Physical Review B}, 107(17):174107, 2023.

\bibitem{wei2022linear}
Xinjian Wei, Congkuan Tian, Hang Cui, Yongkai Li, Shaobo Liu, Ya~Feng, Jian Cui, Yuanjun Song, Zhiwei Wang, and Jian-Hao Chen.
\newblock Linear nonsaturating magnetoresistance in kagome superconductor {Cs}{V}$_3${Sb}$_5$ thin flakes.
\newblock {\em 2D Materials}, 10(1):015010, 2022.

\bibitem{talantsev2015universal}
Evgeny~F Talantsev and Jeffery~L Tallon.
\newblock Universal self-field critical current for thin-film superconductors.
\newblock {\em Nature communications}, 6(1):7820, 2015.

\bibitem{seo2015controlling}
S~Seo, E~Park, ED~Bauer, F~Ronning, JN~Kim, J-H Shim, JD~Thompson, and Tuson Park.
\newblock Controlling superconductivity by tunable quantum critical points.
\newblock {\em Nature Communications}, 6(1):6433, 2015.

\bibitem{ye2024distinct}
Ge~Ye, Mengwei Xie, Chufan Chen, Yanan Zhang, Dongting Zhang, Xin Ma, Xiangyu Zeng, Fanghang Yu, Yi~Liu, Xiaozhi Wang, et~al.
\newblock Distinct pressure evolution of superconductivity and charge density wave in kagome superconductor {Cs}{V}$_3${Sb}$_5$ thin flakes.
\newblock {\em Physical Review B}, 109(5):054501, 2024.

\bibitem{joe2022emergence}
Y.~I. Joe, X.~M. Chen, P.~Ghaemi, K.~D. Finkelstein, G.~A. de~la Pea, Y.~Gan, J.~C.~T. Lee, S.~Yuan, J.~Geck, G.~J. MacDougall, T.~C. Chiang, S.~L. Cooper, E.~Fradkin, and P.~Abbamonte.
\newblock Emergence of charge density wave domain walls above the superconducting dome in ${1T-}\mathrm{TiSe}_{2}$.
\newblock {\em Nature Physics}, 10(6):421--425, 2014.

\bibitem{kim2023monolayer}
Sun-Woo Kim, Hanbit Oh, Eun-Gook Moon, and Youngkuk Kim.
\newblock Monolayer {Kagome} metals {${A}${V}$_3${Sb}$_5$}.
\newblock {\em Nature Communications}, 14(1):591, February 2023.

\bibitem{zhou2023electronic}
Xiaoxiang Zhou, Yongkai Li, Xinwei Fan, Jiahao Hao, Ying Xiang, Zhe Liu, Yaomin Dai, Zhiwei Wang, Yugui Yao, and Hai-Hu Wen.
\newblock Electronic correlations and evolution of the charge density wave in the kagome metals ${A}${V}$_3${Sb}$_5$ (${A}$ = {K, Rb, Cs}).
\newblock {\em Physical Review B}, 107(16):165123, 2023.

\bibitem{yi2023superconducting}
Xin-Wei Yi, Zheng-Wei Liao, Jing-Yang You, Bo~Gu, and Gang Su.
\newblock Superconducting, topological, and transport properties of kagome metals {Cs}{Ti}$_3${Bi}$_5$ and {Rb}{Ti}$_3${Bi}$_5$.
\newblock {\em Research}, 6:0238, 2023.

\bibitem{li2023electronic}
Hong Li, Siyu Cheng, Brenden~R Ortiz, Hengxin Tan, Dominik Werhahn, Keyu Zeng, Dirk Johrendt, Binghai Yan, Ziqiang Wang, Stephen~D Wilson, et~al.
\newblock Electronic nematicity without charge density waves in titanium-based kagome metal.
\newblock {\em Nature Physics}, 19(11):1591--1598, 2023.

\bibitem{labollita2021tuning}
Harrison LaBollita and Antia~S Botana.
\newblock Tuning the van hove singularities in ${A}${V}$_3${Sb}$_5$ (${A}$ = {K, Rb, Cs}) via pressure and doping.
\newblock {\em Physical Review B}, 104(20):205129, 2021.

\end{thebibliography}

\end{document}